\documentclass{article}

\usepackage[main, preprint]{neurips_2026}
\usepackage{amsmath}
\usepackage{algorithm}
\usepackage{algpseudocode}
\usepackage{amssymb}
\usepackage{graphicx}

\usepackage[utf8]{inputenc} 
\usepackage[T1]{fontenc}    
\usepackage{hyperref}       
\usepackage{url}            
\usepackage{booktabs}       
\usepackage{amsfonts}       
\usepackage{nicefrac}       
\usepackage{microtype}      
\usepackage{xcolor}         

\title{TALK: One-Shot Batch Design for Protein Variant Effect Prediction}

\author{%
  \textbf{Qiao Huang}$^{1,2}$
  \qquad
  \textbf{Ao Shen}$^{1,2}$
  \qquad
  \textbf{Jie Du}$^{1,2}$
  \qquad
  \textbf{Manning Wang}$^{1,2}$\thanks{Corresponding author.}
  \\[0.5em]
  $^{1}$ Digital Medical Research Center, School of Basic Medical Sciences, Fudan University, China \\
  $^{2}$ Shanghai Key Laboratory of Medical Image Computing and Computer Assisted Intervention, China \\
  \texttt{\{qahuang24,22111010026,duj23\}@m.fudan.edu.cn}, \;\texttt{mnwang@fudan.edu.cn}
}

\begin{document}

\maketitle

\begin{abstract}
 
Protein variant effect prediction is central to understanding protein function and guiding protein engineering, yet experimental labels remain costly. In many protein experiments, the variants that can be assayed are dictated by protocol constraints, while the variants that need accurate prediction may lie in a different or broader region of the landscape. This assay-prediction mismatch makes candidate-local criteria such as predicted function, uncertainty, or exploration value insufficient for measurement design. Given a protocol-defined accessible space and a limited budget, we ask which variants should be assayed so that the resulting data best improve landscape prediction. We propose TALK, which estimates each candidate’s contribution through posterior coupling between accessible and target variants and reduces redundant information within the batch. Across retrospective ProteinGym and FLIP protocols, TALK generally improves target-space prediction under limited measurement budgets. These results point to a practical two-stage workflow: first, spend a limited assay budget on variants selected to be informative for the target landscape and train a predictive model from the measured labels; then use this model to score and prioritize high-potential variants. Code is available at
\url{https://github.com/joehuang618/TALK}.

\end{abstract}

\section{Introduction}

Accurate protein variant effect prediction is a foundation for understanding protein function and engineering new variants. Wet-lab assays can in principle measure variant function or fitness directly, but systematic enumeration quickly becomes infeasible as mutation order increases. For example, with 50 target positions and 19 non-wild-type substitutions per position, the single-mutant space contains $950$ variants which can still be covered comprehensively by DMS-like assays. The double-mutant space grows to $442{,}225$ variants, and the triple-mutant space to about $1.34\times 10^8$. Reliable experimental characterization requires repeated coverage of each
variant across selection conditions and replicates, making exhaustive
measurement increasingly demanding \citep{fowler2014measuring}. As a result, many experimental systems can obtain relatively systematic labels for singles, but can only measure constrained subsets of higher-order combinatorial spaces due to limits in site choice, library capacity, and screening throughput. Variant effect prediction is valuable in this regime because it learns from limited experimental readouts to predict unmeasured variants and support subsequent candidate prioritization \citep{notin2023proteingym,dallago2021flip}.

Prior work around this prediction-to-design goal has largely followed two related directions. The first is fixed-labeled-set prediction: given existing experimental labels, a model learns sequence-function relationships and predicts unmeasured variants \citep{notin2023proteinnpt,groth2024kermut,teufel2025few}. The second direction integrates predictive models into experimental loops, with machine-learning-guided directed evolution (MLDE) as a prominent example. In MLDE, existing measurements are used to train a surrogate model that predicts variant function. The next experimental batch is then selected from variants predicted to be high-performing by this surrogate, optionally incorporating uncertainty-based exploration terms; after the batch is assayed, the new labels are used to retrain the surrogate \citep{hu2023protein,jiang2024rapid,yang2025active}. With limited initial labels, however, variants predicted to be high-performing by the surrogate can differ substantially from truly high-performing variants, often requiring multiple rounds of experimental feedback to correct the surrogate. Uncertainty-aware criteria can encourage exploration of regions where the model is less certain, but they are still typically defined for the candidate itself or within the current design space. They ask whether a candidate is worth optimizing or exploring, not whether measuring it would improve posterior prediction on a broader landscape. Consequently, costly multi-round iteration may not identify high-performing variants early, and it does not explicitly ensure that each measurement is most useful for improving prediction.

We therefore focus on a structural mismatch in candidate selection: the variants that a protocol allows us to measure need not be the variants on which we ultimately want better predictions. 

This differs from standard pool-based active learning, where the variants selected for measurement and the variants used to evaluate prediction improvement are usually drawn from the same space. In protein variant experiments, these two sets can differ: a protocol may allow only certain variants to be assayed, while the variants that matter for prediction or engineering may lie in a different part of the landscape. For example, the measurable candidates may be single mutants or variants constrained by site choice, synthesis cost, and screening throughput. The target space may instead be double mutants, higher-order combinations, or a prespecified panel obtained from functional goals, structural constraints, or computational screening rules. A useful feature of this setting is that the target space can be specified before selection, even though their functional labels are not known. Selection therefore cannot use target labels, but a predictive model can use the posterior relationship between measurable sequences and target sequences to estimate which measurements would most improve prediction on the target space. Therefore, selecting variants only by their predicted function or uncertainty within the accessible space is insufficient. A variant that appears high-performing or highly uncertain among single mutants may not be the measurement that most improves prediction for double mutants or higher-order combinations. The relevant question is whether measuring it would change the model's posterior judgment on the target space. We formalize this structure by denoting the protocol-defined measurable space as the accessible space $S$ and the final space for prediction as the target space $A$. The two sets may be disjoint, partially overlapping, nested in either direction, or identical. The key constraint is that selection may use sequences in the target space to decide which variants in the accessible space to measure, but it must not use labels from the target space.

This structure, measuring one set to improve prediction on another known set, can be viewed as a transductive active learning problem: accessible samples and prediction targets may belong to different sets, and selection value is defined by benefit to the target set \citep{hubotter2024transductive}. Building on this view, we propose TALK, a batch selection framework for protein variant experimental design. TALK first uses protein-native posterior coupling to estimate how measuring a variant in the accessible space would affect posterior prediction on the target space. It then scores candidate measurements by their contribution to target space posterior prediction. Finally, TALK turns single-point measurement value into a one-shot batch selection rule by applying covariance-only conditioning, reducing redundant measurements within the batch. The resulting batch is therefore directed toward improving prediction.

Our contributions are as follows:

\begin{itemize}
\item We formulate protein variant selection as a batch design problem: the experimenter selects a limited number of variants from the protocol-defined accessible space to improve posterior prediction on the target space.

\item We introduce TALK, a batch selection method that selects variants from the accessible space to improve posterior prediction on the target space with a single planned batch, without requiring multiple rounds of label feedback. TALK estimates each variant's contribution from protein-native posterior coupling and uses conditional batch construction to reduce redundancy within the batch.

\item We evaluate TALK on retrospective ProteinGym and FLIP protocols, demonstrating improved target-space prediction under limited measurement
budgets. Strong-baseline comparisons and sensitivity analyses characterize
its performance and scope.
\end{itemize}

\section{Related Work}

Protein variant effect prediction has developed along two relevant lines. On the prediction side, benchmarks such as FLIP and ProteinGym have enabled systematic evaluation of variant-effect predictors \citep{dallago2021flip,notin2023proteingym}, and recent models improve extrapolation using contextual labels, evolutionary or language-model priors, retrieval, and protein-native Bayesian structure \citep{notin2023proteinnpt,teufel2025few,groth2024kermut,frazer2021disease,notin2022tranception,truong2023poet}. On the design side, machine-learning-guided directed evolution (MLDE) methods such as BO-EVO, EVOLVEpro, and ALDE \citep{biswas2021low,hu2023protein,jiang2024rapid,yang2025active} use model scores, uncertainty, low-order mutation information, or exploratory criteria to nominate variants for subsequent experiments. MULTI-evolve \citep{tran2026rapid} further shows that low-order mutations and epistasis information can support higher-order multi-mutant design. TALK asks how a measurement budget should be allocated within accessible space to improve prediction on the target space.

Active learning and Bayesian experimental design provide a general language for selecting valuable measurements \citep{mackay1992information,chaloner1995bayesian,cohn1996active}, including information-gain, submodular, and batch-selection objectives \citep{krause2012near,nemhauser1978analysis,kirsch2019batchbald,ash2019deep}. Most such objectives are defined with respect to a pool or overall model learning. The closest connection to TALK is transductive active learning, where selection value is defined by its effect on specified prediction targets and the accessible space may differ from the target space \citep{seo2000gaussian,yu2006active,fiez2019sequential,smith2023prediction,hubotter2024transductive}. TALK adapts this view to protein variant experiments, where experimental accessibility, targets, and one-shot batch budgets are determined by the experimental protocol.

\section{Method}
We formulate protein variant selection as a budget design problem, where a limited batch is selected from the protocol-defined accessible space to improve prediction on the target space. Section~\ref{31}  formally defines the accessible space $S$, the target space $A$, and the fixed-budget batch objective. Section~\ref{32} then specifies the posterior interface used to relate measurable candidates to target variants. Section~\ref{33} introduces two acquisition scores from this candidate-target posterior coupling, and Section~\ref{34} extends single-candidate scores to non-redundant one-shot batch selection. Finally, Section~\ref{35} gives the resulting algorithm and instantiates concrete protein experimental settings.

\subsection{Problem Setup}
\label{31}
Let $X$ denote the finite set of protein variants for a given assay, and let $f:X\to\mathbb{R}$ be the unknown variant-effect function. Measuring a variant $x\in X$ yields a noisy label $y_x=f(x)+\epsilon_x$, where $\epsilon_x$ is zero-mean observation noise with variance $\tau^2$. At acquisition step $n$, the observed dataset is $\mathcal{D}_n=\{(x_i,y_i)\}_{i=1}^{n}$. We write $L_n=\{x_i:(x_i,y_i)\in\mathcal{D}_n\}$ for the set of labeled variants and $y_n=(y_1,\ldots,y_n)^\top$ for the corresponding label vector.

We distinguish two subsets of $X$. The accessible space $S\subseteq X$ is the set of variants that can be selected and measured under the current experimental protocol. The target space $A\subseteq X$ is the set of variants relevant for final evaluation or decision making. The relationship between $S$ and $A$ may be disjoint, partially overlapping, nested in either direction, or identical. If a target variant also belongs to $S$, its label enters $\mathcal{D}_n$ only if the variant is actually selected. The unmeasured subset of the accessible space at step $n$ is $U_n=S\setminus L_n$. Given a labeling budget $B$, we seek a batch $Q_n\subseteq U_n$ with $|Q_n|=B$ whose measurements maximally improve prediction on $A$.

This $S\to A$ formulation follows the distinction between sample space and target space in transductive active learning. In our setting $S$ and $A$ are jointly determined by experimental protocols, mutation order, target panels, and evaluation boundaries. This differs from pool-internal active learning \citep{cohn1996active} or standard sample-space Bayesian optimization \citep{srinivas2009gaussian}. In those settings, a variant is valuable because it is more uncertain, more diverse, or more promising within the pool. Here, a variant is worth measuring if observing it would change posterior beliefs over the target space.

\subsection{Posterior Interface}
\label{32}

TALK uses Kermut as a protein-native Bayesian surrogate \citep{groth2024kermut}. Kermut is a Gaussian-process model whose protein-informed kernel defines posterior dependence among variants. Given the sequences of $A$ and $U_n$ and the labels observed in $L_n$, Kermut GP conditioning produces predictive means $\mu_n(A)$, the posterior covariance over targets $\Sigma_n(A,A)$, and candidate-target cross-covariances $\Sigma_n(A,U_n)$. These are exactly the quantities needed by the experimental-design interface.

Although our experiments use Kermut to instantiate this interface, TALK is not tied to Kermut: its acquisition rules only require finite-set posterior means and covariances. We write the surrogate interface in Gaussian-process notation \citep{seeger2004gaussian}. Let $f\sim\operatorname{GP}(m,k)$, where $m$ is a mean function and $k$ is a positive semidefinite covariance function over protein variants. Given the current observations $\mathcal{D}_n$, for any finite variant sets $C,C'\subseteq X$, denote the posterior mean and covariance by:

\begin{align}
\mu_n(C)
&=m_C+K_{C,L_n}(K_{L_n,L_n}+\tau^2 I)^{-1}(y_n-m_{L_n}),\\
\Sigma_n(C,C')
&=K_{C,C'}-K_{C,L_n}(K_{L_n,L_n}+\tau^2 I)^{-1}K_{L_n,C'}.
\end{align}

Here $K_{C,C'}$ denotes the covariance matrix induced by the kernel $k$ on sets $C$ and $C'$. For prediction, the posterior mean $\mu_n(A)$ provides target-set estimates, and $\operatorname{diag}\Sigma_n(A,A)$ provides marginal predictive uncertainty. For acquisition, TALK uses the cross-covariance $\Sigma_n(A,U_n)$ between target variants and unmeasured accessible variants. This cross-covariance converts a surrogate into measurement value: it determines how observing a candidate in $S$ is expected to change beliefs over $A$. With model parameters fixed, $\mu_n(A)$ depends on both $L_n$ and $y_n$,
whereas $\Sigma_n(A,U_n)$ depends on $L_n$ but not $y_n$, enabling
label-free covariance calculation during batch selection. Appendices~\ref{appc}, \ref{appd}, and~\ref{appe} describe the large-target posterior construction, the interface and complexity.

\subsection{Acquisition Scores}
\label{33}

Given $\Sigma_n(A,A)$ and $\Sigma_n(A,U_n)$, TALK scores each $x\in U_n$ by the posterior change that a noisy observation $y_x$ would induce on $f_A$. We introduce two scores from \citep{hubotter2024transductive}: Information-based Transductive Learning (ITL) for information gain and Variance-based Transductive Learning (VTL) for variance reduction. ITL and VTL define the value of a candidate by its posterior effect on the target set in two ways. ITL selects the candidate whose observation gives the largest information gain about the target effects. VTL selects the candidate that would produce the largest total marginal variance reduction over $A$. Their single-candidate selection rules are:

\begin{align}
x_{n+1}^{\mathrm{ITL}}
&=\arg\max_{x\in U_n} I(f_A;y_x\mid\mathcal{D}_n), \\
x_{n+1}^{\mathrm{VTL}}
&=
\arg\min_{x\in U_n}
\operatorname{tr}
\operatorname{Var}\left[
f_A\mid\mathcal{D}_n,y_x
\right].
\end{align}

Under a Gaussian posterior, the corresponding covariance-based scores can be written as:

\begin{gather}
s_{\mathrm{ITL}}(x;A)
=-\frac{1}{2}
\log\left(
1-\frac{\Sigma_n(x,A)\Sigma_n(A,A)^{-1}\Sigma_n(A,x)}
{\Sigma_n(x,x)+\tau^2}
\right), \\
s_{\mathrm{VTL}}(x;A)
=\sum_{a\in A}
\frac{\Sigma_n(a,x)^2}{\Sigma_n(x,x)+\tau^2}.
\end{gather}

For both ITL and VTL, larger scores correspond to more valuable measurements, so candidates are selected by maximizing the score over the current measurable pool. The denominator $\Sigma_n(x,x)+\tau^2$ is the total observation variance of candidate $x$, including observation noise, and the numerator terms quantify how strongly $x$ is posterior-coupled to the target variants. Following Kermut, the observation noise $\tau^2$ is learned jointly
with the kernel hyperparameters by optimizing the exact marginal likelihood
with a Half-Cauchy prior of scale 0.1. ITL evaluates this coupling at the level of the full target vector through $\Sigma_n(A,A)^{-1}$, thereby accounting for covariance structure within $A$. VTL sums the marginal variance reductions over target variants without inverting $\Sigma_n(A,A)$, making it computationally more efficient for large target sets.

\subsection{Batch Selection}
\label{34}

In practice, variants in the same experimental round are synthesized and assayed together, so within-batch labels are unavailable at selection time. Single-candidate scores must therefore be converted into a non-redundant batch that avoids repeatedly measuring the same posterior uncertainty directions.

Ideally, the batch should directly optimize its overall posterior benefit for $A$, which can be written as a non-adaptive combinatorial design problem:

\begin{equation}
\vspace{-1em}
Q_n^\star=\arg\max_{Q\subseteq U_n,\ |Q|=B}G_n(Q;A),
\end{equation}

where $G_n(Q;A)$ is the overall acquisition score of batch $Q$ for target set $A$. Solving this objective exactly would require comparing $\binom{|U_n|}{B}$ batches, which is infeasible at realistic protein-library scales. We therefore use conditional greedy batch construction as an executable approximation. After each selected variant, TALK performs covariance-only conditioning and recomputes the marginal gain of the remaining candidates with respect to $A$. This makes later choices valuable only if they reduce target uncertainty not already covered by the current batch.

\begin{algorithm}[t]
\caption{TALK conditional one-shot batch selection}
\label{alg:talk}
\begin{algorithmic}[1]
\Statex \textbf{Input:} Initial labels $\mathcal{D}_0$; accessible space $S$; target space $A$; budget $B$; observation noise $\tau^2$; acquisition rule $g\in\{\mathrm{ITL},\mathrm{VTL}\}$
\Statex \textbf{Output:} Selected batch $Q$
\State Construct labeled set $L_0 \leftarrow \{x:(x,y)\in\mathcal{D}_0\}$
\State Define candidate pool $U_0 \leftarrow S\setminus L_0$
\State Fit the Kermut posterior using $\mathcal{D}_0$
\State Build posterior objects over $U_0$ and $A$ required by acquisition rule $g$
\State Initialize selected batch $Q \leftarrow \varnothing$
\State Let $\Sigma^{(0)}$ be the posterior covariance over $A\cup U_0$
\For{$j=1,\ldots,B$}
    \For{each $x\in U_0\setminus Q$}
        \State Compute marginal gain $s_g^{(j)}(x;A)$ from $\Sigma^{(j-1)}$
    \EndFor
    \State Select $x_j \leftarrow \arg\max_{x\in U_0\setminus Q} s_g^{(j)}(x;A)$
    \State Update $Q \leftarrow Q\cup\{x_j\}$
    \State Update $\Sigma^{(j)}$ by covariance-only Gaussian conditioning on $x_j$
\EndFor
\State \Return $Q$
\end{algorithmic}
\end{algorithm}

Let $Q_j=\{x_1,\ldots,x_j\}$ denote the first $j$ variants selected into the batch, with $Q_0=\varnothing$. Let $\Sigma^{(0)}=\Sigma_n$ be the posterior covariance before batch construction. At greedy step $j$, TALK uses the covariance $\Sigma^{(j-1)}$ conditioned on $Q_{j-1}$ and selects:

\begin{equation}
\vspace{-1em}
x_j
=
\arg\max_{x\in U_n\setminus Q_{j-1}}
s^{(j-1)}(x),
\end{equation}

where $s^{(j-1)}(x)$ is defined by the corresponding acquisition score, computed using the current conditional covariance $\Sigma^{(j-1)}$. After selecting $x_j$, we set $Q_j=Q_{j-1}\cup\{x_j\}$ and update the covariance as:

\begin{equation}
\Sigma^{(j)}(C,C')
=
\Sigma^{(j-1)}(C,C')
-
\frac{
\Sigma^{(j-1)}(C,x_j)\Sigma^{(j-1)}(x_j,C')
}{
\Sigma^{(j-1)}(x_j,x_j)+\tau^2
}.
\end{equation}

Here $C,C'\subseteq X$ are arbitrary finite sets. This can be used to update $\Sigma^{(j)}(A,A)$ and  $\Sigma^{(j)}(A,U_n\setminus Q_j)$ for the next selection. The update depends only on posterior covariance and observation noise, not on the label $y_{x_j}$. Appendix~\ref{appa} gives the Gaussian conditioning form used in computation. Appendix~\ref{appb} states the  conditions for greedy selection guarantees.

\subsection{Protocol Instantiations and Execution}
\label{35}
The posterior interface, acquisition scores, and conditional batch rule define TALK's computational core. To apply them to a concrete protein experiment, the protocol specifies three sets: the initial labeled set $L_0$, the accessible space $S$ from which additional measurements may be selected, and the target space $A$ of final interest. TALK then fits a posterior from the available labels and selects a batch of size $B$ from $U_0=S\setminus L_0$. Algorithm~\ref{alg:talk} summarizes the conditional construction of a one-shot batch. TALK first fits the Kermut posterior using the currently available labels, constructs the posterior mean and covariance objects over $S$ and $A$, and then greedily selects a size-$B$ batch while updating covariance to avoid redundant selections.

TALK plans a non-redundant batch before true labels return, turning a fixed wet-lab budget into an actionable set of measurements. TALK supports both one-shot experiments and iterative multi-round workflows. If an adequately sized one-shot batch already reaches the desired performance, the experimenter need not repeatedly wait for feedback to revise selections round by round, reducing delays, operational cost, and batch-condition variation in multi-round experiments. If multiple rounds are still needed, the newly observed labels can be added to $\mathcal{D}$ after each round, the posterior can be refit, and the same batch planner can be run again.

\section{Experiments}
\subsection{Experimental Setup}
\label{41}

The experiments test TALK's central claim: under a fixed budget, measurements should be selected from the accessible space $S$ according to their expected contribution to prediction on the target space $A$. We evaluate this claim in various settings. We refer to the 51-assay panel selected by \citep{groth2024kermut} from ProteinGym assays containing multi-mutant variants as ProteinGym-MM; this panel supports multi-mutation-order transfer protocols across assays. The GB1 complete landscape from FLIP \citep{dallago2021flip} provides a complete combinatorial fitness landscape for evaluating higher-order target prediction without additional ambiguity from biased subsampling.

We describe all protocols using the $L_0,S,A$ notation. We use $kM$ for variants with $k$ substitutions relative to the wild type, e.g., $1M$ and $2M$ for single and double mutants. Section \ref{42} evaluates ProteinGym-MM under canonical $S=1M\to A=2M$, a mixed-order pool protocol $S=1M/2M\to A=2M$, and a singles-premeasured protocol $L_0=1M, S=2M\to A=2M$. Section \ref{43} evaluates GB1 complete-landscape order-to-rest transfer, where low-order accessible spaces serve higher-order targets. Section~\ref{44} compares conditional one-shot selection with a
no-conditioning control and directed-evolution-style iterative selection with multi-round label feedback.

The primary metric is Spearman's rank correlation coefficient on the test set, which measures how acquisition improves the ranking of target variants. We also report RMSE and top-5\% target recovery as supplementary metrics. We report a static reference for each setting, defined as the protocol-matched performance obtained by Kermut after using all labels in the accessible space. This is an empirical protocol-matched reference, not a mathematical upper bound. We compare ITL and VTL against Random, SampleVariance, top-mean, and UCB, corresponding to random, marginal uncertainty, predictive mean, and mean-plus-uncertainty, respectively \citep{greenman2025benchmarking,jiang2024rapid,srinivas2009gaussian}. For Random, we report the average performance over five seeds. Full results are reported in Appendix~\ref{apph}; budget coverage and sample-saving efficiency are reported in Appendix~\ref{appg}; $S=1M/2M \to A=1M $ and $S=1M/2M\to A\geq3M$ protocols are reported in Appendix~\ref{appi}. Appendix~\ref{app:target-uncertainty-diagnostics} further shows performance gains are accompanied by faster reduction in target uncertainty.

\begin{figure}[t]
\centering
\includegraphics[width=0.9\linewidth]{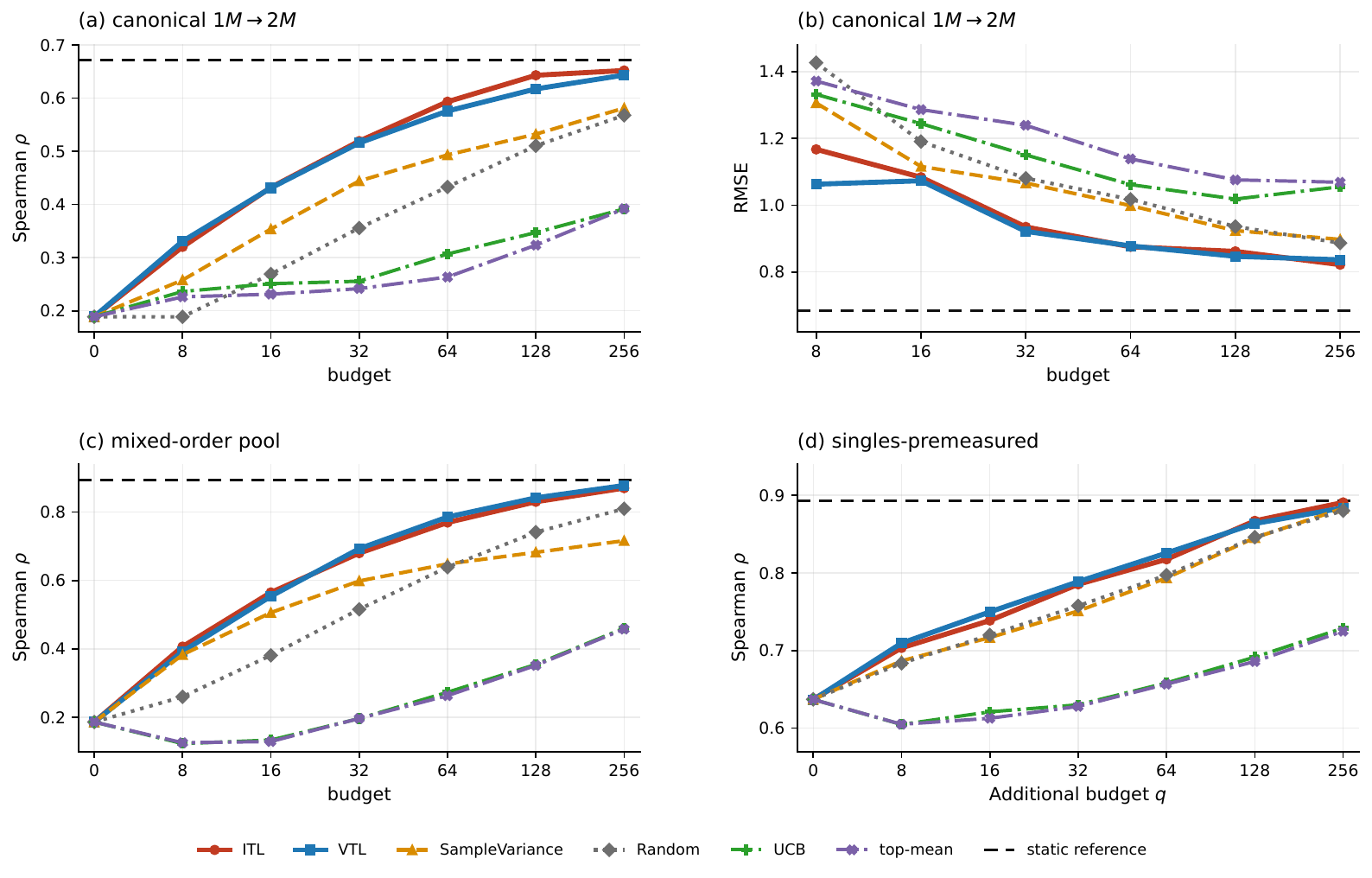}
\caption{Target prediction across three ProteinGym-MM protocols. Panels
(a,b) show canonical cold-start $S=1M\to A=2M$ Spearman and RMSE; panels (c,d) show
mixed-order and singles-premeasured Spearman. Dashed static references are empirical full-pool fits.}
\label{fig1:pgmulti51-protocols}
\end{figure}

\subsection{Mutation-Order Transfer on ProteinGym-MM}
\label{42}

The first experiments study mutation-order transfer, where lower-order or otherwise accessible variants are measured to improve prediction in a different mutation space. We first test direct $1M\to2M$ transfer, then use fold-based protocols to assess whether the same principle holds under benchmark-style and wet-lab-like splits.

\subsubsection{Canonical $S=1M\to A=2M$}
In the canonical $1M\to2M$ setting, the accessible space $S$ is the set of all $1M$ variants in the assay, and the target space $A$ is the set of all $2M$ variants. A method may select only fixed-budget one-shot measurements from $S=1M$, while evaluation is performed on $A=2M$.

The panels (a,b) of Figure~\ref{fig1:pgmulti51-protocols} show the canonical $1M\to2M$ transfer results in terms of Spearman $\rho$ and RMSE. ITL and VTL improve the target-space ranking rapidly at small budgets and approach the full-$1M$ static reference by $B=128$. At $B=64$, ITL already exceeds the $B=256$ Spearman performance of Random and SampleVariance; VTL is close to SampleVariance at $B=256$ and remains ahead of both baselines in RMSE. The panel (b) shows the same pattern from the error perspective: ITL/VTL reduce target prediction error much faster, indicating that the gain is not limited to rank correlation. Thus, in the canonical $1M\to2M$ setting, TALK achieves comparable or better target prediction than 256-budget baselines using roughly one quarter of the measurement budget, corresponding to about 75\% label saving in this comparison. Top-mean and UCB perform poorly in target-space ranking, further supporting
the need to select measurements for their contribution to target prediction
rather than candidate-local value.

\subsubsection{Mixed-Order Pool and Singles-Premeasured Protocols }

To test whether TALK transfers to settings closer to wet-lab workflows, we further construct mixed-order protocols using the \texttt{fold\_rand\_multiples} split scheme in ProteinGym. This split partitions multi-mutant variants into train-side and target-side sets. We fix the target space as $A=2M_{\mathrm{test}}$, and acquisition selects measurements from train-side $1M/2M$ variants that are most useful for predicting target-side double mutants.

As shown in the panel (c) of Figure~\ref{fig1:pgmulti51-protocols}, in the mixed-order pool protocol, the clearest difference is early-budget sample efficiency. By $B=32$, ITL and VTL already raise Spearman $\rho$ to about $0.68$-$0.69$, close to the level that SampleVariance reaches only near $B=256$. Thus, in this setting, posterior-coupled acquisition shifts the budget-performance curve substantially leftward: a small fraction of the train-side pool captures much of the target-side predictive gain that less targeted rules obtain only at later budgets. As the budget increases, ITL/VTL continue moving toward the static reference, suggesting that the early advantage reflects a more efficient ordering of informative measurements rather than a transient low-budget effect.

The singles-premeasured protocol provides an additional view. Once all $1M$ variants are included in $L_0$, the posterior starts from a stronger estimate of $2M_{\mathrm{test}}$, and the problem becomes double-mutant refinement. As shown in the panel (d) of Figure~\ref{fig1:pgmulti51-protocols}, even from this stronger starting point, ITL/VTL move toward the static reference more quickly than other baselines. Thus, a complete singles scan improves the initial posterior, but limited follow-up double-mutant measurements still benefit from acquisition rules that prioritize target-side posterior change.

\begin{figure}[t]
\centering
\includegraphics[width=\linewidth]{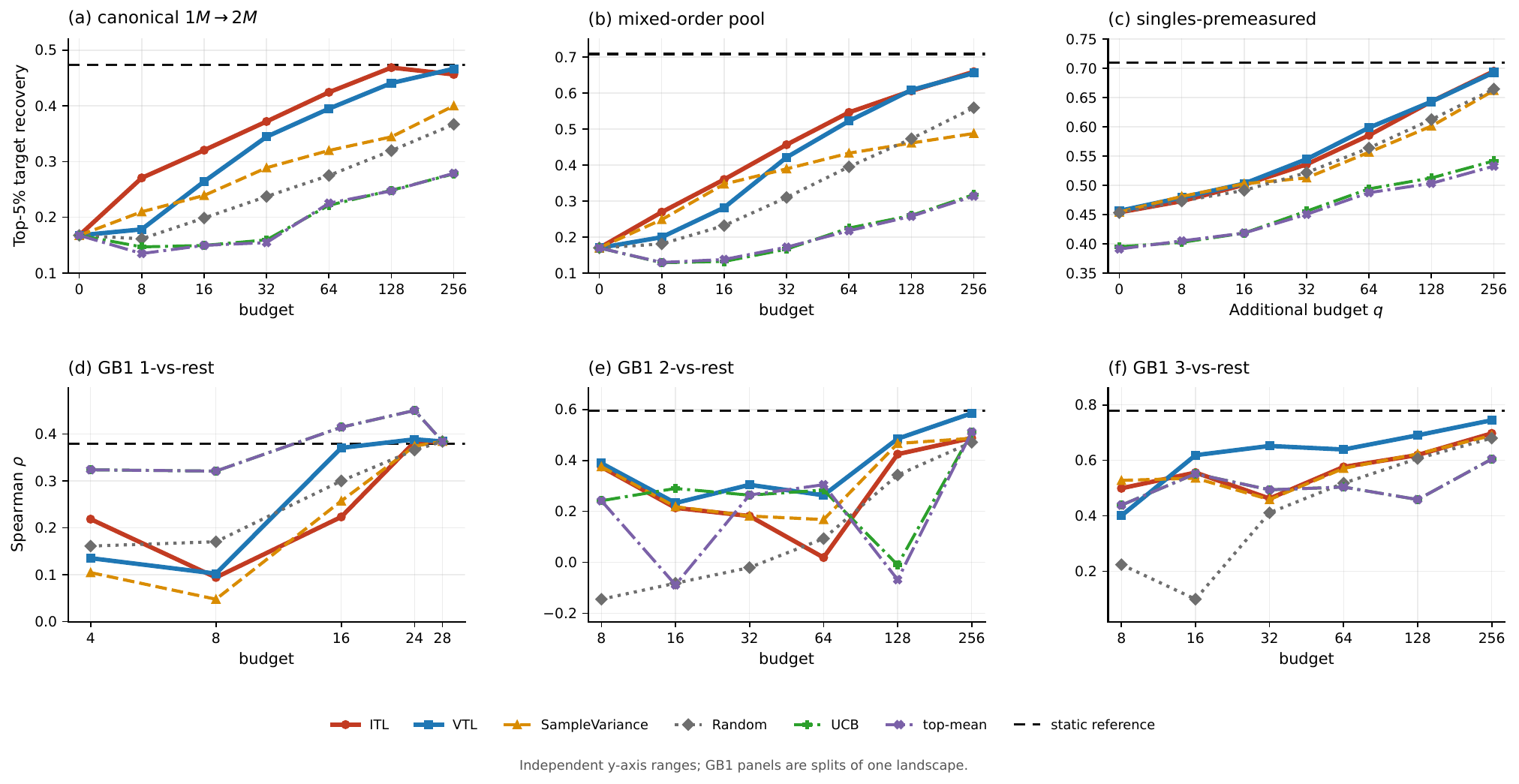}
\caption{Top-5\% recovery on three ProteinGym-MM protocols and Spearman
correlation on three GB1 order-to-rest protocols. The top row reports top-5\% target recovery for canonical $1M\to2M$, the mixed-order pool protocol, and the singles-premeasured protocol. The bottom row reports Spearman $\rho$ on the GB1 complete landscape for \texttt{1-vs-rest}, \texttt{2-vs-rest}, and \texttt{3-vs-rest}.}
\label{fig2:pgmulti51-top5-gb1}
\end{figure}

\subsubsection{Top-5\% Target Recovery Across Three Mutation-Space Settings}

We also evaluate top-5\% target recovery, defined as the overlap between the predicted and true top tails, across the three protocols. This metric emphasizes the variants most relevant for experimental follow-up. As shown in the top three panels in Figure~\ref{fig2:pgmulti51-top5-gb1}, ITL/VTL remain strongest or near-strongest across the three protocols, with clear gains over Random in $1M\to2M$ and the mixed-order pool protocol and competitive performance in the singles-premeasured setting. The ranking gains from acquisition also translate to better recovery of the most valuable region of the target space.

\subsection{GB1 Complete-Landscape Evaluation}
\label{43}
GB1 provides a complete combinatorial landscape over four target positions in a single protein, enabling a full-space test of TALK's target transfer across mutation orders. Because all combinations over the four sites are observed in this landscape, we can directly construct order-to-rest settings from lower-order accessible spaces to higher-order target spaces, avoiding additional ambiguity from incomplete sampling.

We use the semantically clearest \texttt{1/2/3-vs-rest} tasks from FLIP and directly inherit their train/test splits. For each task, a method may select one batch of labels only from the train-side accessible space $S$, and Spearman rank correlation is evaluated on the target space $A$. \texttt{1-vs-rest} corresponds to $S=1M$ and $A=2M\cup 3M\cup 4M$; \texttt{2-vs-rest} corresponds to $S=1M\cup 2M$ and $A=3M\cup 4M$; and \texttt{3-vs-rest} corresponds to $S=1M\cup 2M\cup 3M$ and $A=4M$.

In the \texttt{1-vs-rest} setting shown in Figure~\ref{fig2:pgmulti51-top5-gb1}(d), the accessible pool contains only \texttt{WT\ +\ 27} single mutants, so a few high-impact variants can cause substantial local fluctuations. In this setting, top-mean and UCB may exceed the static reference at some budgets; this reflects fluctuations between an empirical reference curve and near-exhaustive subsets, rather than showing that exploitation generally solves higher-order transfer. At $B=28$, all methods use the full measurable pool and converge to the same result. Thus, \texttt{1-vs-rest} is better interpreted as a small-pool boundary case.

The more discriminative GB1 tasks are \texttt{2-vs-rest} and \texttt{3-vs-rest}, where lower-order measurements must support prediction on triple/quadruple targets. Across these two panels, VTL is the most stable method and stays closest to the static reference. VTL exceeds Random and conditional SampleVariance at the reported budgets, reaching Spearman 0.5851 versus 0.4712 and 0.4868 at $B=256$. This is consistent with its trace-variance objective: VTL aggregates candidate-target posterior coupling across target variants, favoring measurements that broadly reduce uncertainty over the higher-order space. GB1 therefore provides a complete-landscape check showing that target-side variance reduction can efficiently convert limited low-order measurements into higher-order prediction gains.

\subsection{Iterative Directed-Evolution-Style Comparison}
\label{44}

\begin{figure}[t]
\centering
\includegraphics[width=\linewidth]{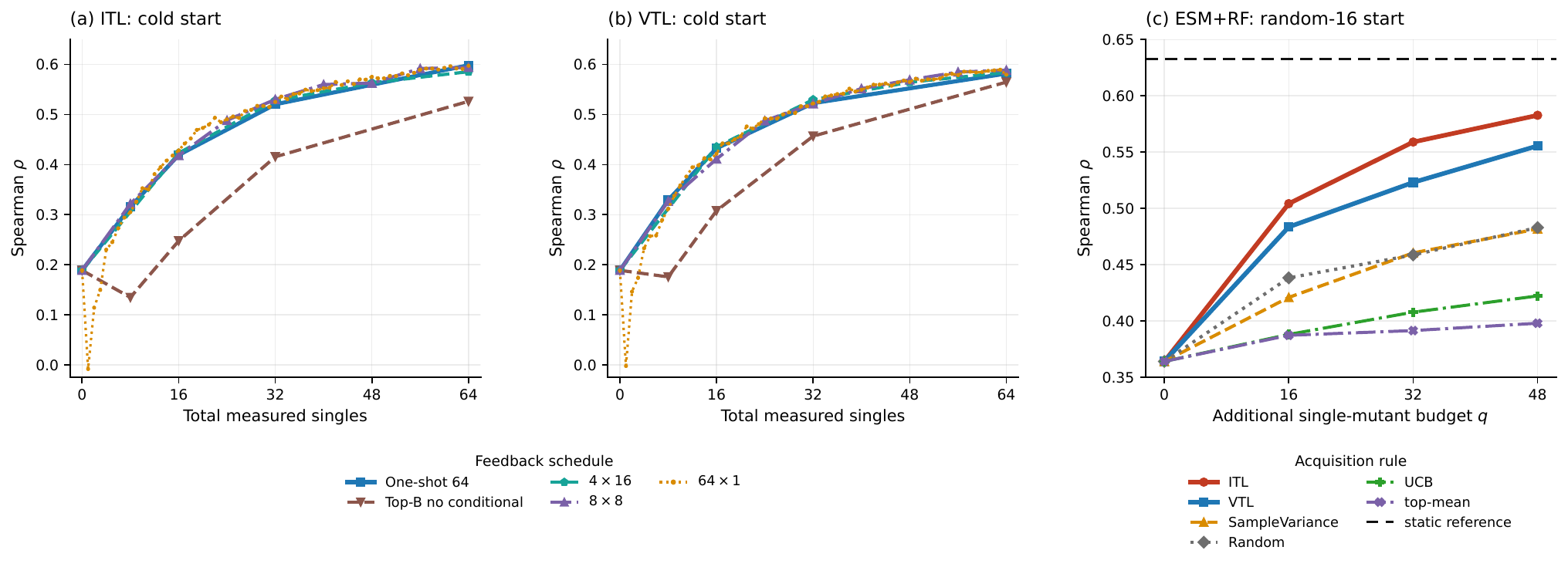}
\caption{Feedback schedules and a separate prediction-interface check.
Panels (a,b) use canonical $1M\to2M$ cold start and a total budget of 64
measured singles, comparing conditional one-shot selection, unconditioned
top-$B$ selection, and feedback schedules. Panel (c) fixes an ESM+RF final
predictor and starts from 16 randomly measured singles; its x-axis counts
additional measurements.}
\label{fig:p51-iterative-uncertainty}
\end{figure}

This auxiliary comparison connects TALK to directed-evolution-style model-in-the-loop settings. We fix the selector family and canonical $1M\to2M$ protocol under cold start, with $L_0=\varnothing$, $S=1M$, $A=2M$, and a total budget of 64 measured singles. The comparison isolates two design choices: whether the batch uses covariance conditioning, and whether true labels are returned between rounds.

Panels (a) and (b) of Figure~\ref{fig:p51-iterative-uncertainty} show that conditional batch construction is the dominant factor. Feedback does not show a monotonic
advantage, although it may help under other posterior or protocol conditions. No-conditional top-$B$ uses the same single-point ITL/VTL scores but ranks candidates only once under the initial posterior. Conditional one-shot selection instead updates the covariance after each selected candidate and recomputes marginal gains, reducing redundancy among measurements.

Finer feedback schedules change the selected variants but do not yield a monotonic advantage. In \texttt{4x16}, \texttt{8x8}, and \texttt{64x1}, returned labels are added after each round, the posterior is refit, and the next batch is planned from the updated posterior; nevertheless, performance remains close to one-shot. The final selected sets also retain substantial overlap with one-shot 64: 73.76\%, 72.82\%, and 71.85\% for ITL, and 64.74\%, 62.88\%, and 59.87\% for VTL across \texttt{4x16}, \texttt{8x8}, and \texttt{64x1}. This comparison isolates the benefit of conditional updates in constructing an informative, low-redundancy batch.

These results indicate that conditional one-shot planning already captures many measurements that remain valuable under alternative feedback schedules. Multi-round feedback is still useful for correcting the posterior after labels return, but in this setting it does not substitute for non-redundant batch construction. This suggests a practical two-stage workflow with TALK: first use an adequately sized one-shot batch to map target-relevant uncertainty, and then use the updated posterior to focus a subsequent round on the most promising regions or candidate variants.

\section{Discussion and Limitations}

TALK reallocates a fixed measurement budget toward the target-relevant information available through the accessible space $S$. It should therefore be interpreted as a way to approach the protocol-matched static reference more efficiently, rather than as a mechanism for systematically exceeding it: the static reference fits the surrogate using all labels in the accessible training pool. The remaining gains and saturation behavior depend on how strongly $S$ supports $A$ under the selection posterior. When $\Sigma(A,S)$ is strong, a small number of measurements in $S$ can substantially improve target-space prediction; when candidate-target coupling is weak, or when the posterior misses relevant structural, site-combination, or epistatic structure, acquisition curves will saturate regardless of the selection. Before selection, practitioners can compare candidate pools using normalized posterior-reducible target variance, together with budget-aware and
model-sensitivity checks (Appendix~\ref{appf}). These diagnostics describe
support under the declared posterior rather than a universal success
threshold.

Crucially, this support condition is an interface-level requirement rather than a commitment to one fixed model class. TALK requires a posterior or covariance-like selection interface, not a specific Kermut predictor. ITL/VTL scores and conditional batching depend only on finite-set posterior quantities such as $\Sigma_n(A,A)$, $\Sigma_n(x,x)$, $\tau^2$, and $\Sigma_n(A,x)$. Thus, a Kermut GP, an embedding-based Gaussian posterior, a Bayesian last-layer model, or another covariance-like surrogate can instantiate TALK when it provides reliable coupling for the current $S/A$ protocol.

Figure~\ref{fig:p51-iterative-uncertainty}(c) provides an auxiliary check of this interface view. With ESM+RF fixed as the final predictor, ESM+RF + ITL/VTL outperform top-mean, UCB, Random, and SampleVariance under the same $1M\to2M$ protocol with 16 random variants warm-start. This suggests that the gains arise from target-coupled measurement selection rather than from Kermut alone, the number of added labels, pointwise uncertainty, or sample-space UCB. Overall, TALK should be understood as a protocol-aware measurement-design framework with interface-level flexibility: reliable candidate-target coupling allows the selection stage to translate limited wet-lab budgets more directly into target-space prediction gains.

In this work, all evaluations are retrospective; prospective wet-lab validation remains
future work. Gains depend on accessible--target support and posterior
fidelity. When approximate posterior backends or restricted target panels are used, the acquisition scores reflect the specified approximations rather than exact full-space objectives.

\section{Conclusion}

We presented TALK, a transductive active learning framework for protein variant measurement design. TALK selects measurements according to their expected contribution to target-space prediction rather than their own predicted value or uncertainty. This reframes protein variant acquisition as a question of what to measure first in order to learn the landscape that matters.

Looking forward, this view suggests a more direct workflow for protein engineering: use an adequate, carefully planned one-shot experiment not merely to search for immediate hits, but to probe the landscape efficiently before larger design decisions are made. Such batches can offer a fast first map of difficult regions of the landscape and guide subsequent experimental decisions efficiently. As protein design moves deeper into combinatorial spaces, measurement
design may become as important as model design itself, helping both budget-limited and high-throughput experiments turn finite measurement capacity into compact, actionable maps for discovery.

\begin{ack}
This work was supported by the Science and Technology Innovation Plan of Shanghai Science and Technology Commission under Grant 24J22800800 and Fudan University Science Intelligence Special Fund under Grant FD-AI4S04183. Additional support was provided by the CFFF Intelligent Computing Platform at Fudan University.
\end{ack}

\bibliographystyle{plainnat}
\bibliography{references}


\newpage

\appendix
\counterwithin{figure}{section}
\counterwithin{table}{section}

\section{Gaussian Decision Rules Used by TALK}
\label{appa}

This appendix gives the finite-dimensional Gaussian decision rules used by TALK \citep{hubotter2024transductive,seeger2004gaussian}, including posterior quantities, single-point acquisition scores, and batch covariance updates. At the start of an acquisition round, TALK uses the finite marginal of the current posterior over the target panel and unqueried candidates:

\begin{equation}
\begin{bmatrix}
f_A\\
f_{U_n}
\end{bmatrix}
\mid \mathcal{D}_n
\sim
\mathcal{N}\!\left(
\begin{bmatrix}
\mu_n(A)\\
\mu_n(U_n)
\end{bmatrix},
\begin{bmatrix}
\Sigma_n(A,A) & \Sigma_n(A,U_n)\\
\Sigma_n(U_n,A) & \Sigma_n(U_n,U_n)
\end{bmatrix}
\right).
\end{equation}

If $x\in U_n$ is queried, the observation is $y_x=f_x+\epsilon_x$, with $\epsilon_x\sim\mathcal{N}(0,\tau^2)$. In batch planning, after conditioning on a location $x$ that will be measured, the posterior covariance update is

\begin{equation}
\Sigma_n^+(C,C')
=
\Sigma_n(C,C')
-
\frac{
\Sigma_n(C,x)\Sigma_n(x,C')
}{
\Sigma_n(x,x)+\tau^2
}.
\end{equation}

This update does not depend on the realized label value, so covariance can be used to plan subsequent candidates within a one-shot wet-lab batch. The true labels enter $\mathcal{D}_n$ only when the posterior is refit in the next round.

Under the same finite Gaussian posterior, the single-point VTL score is the reduction in trace variance on $A$ after measuring $x$:

\begin{equation}
s_{\mathrm{VTL}}(x;A)
=
\sum_{a\in A}
\frac{
\Sigma_n(a,x)^2
}{
\Sigma_n(x,x)+\tau^2
}.
\end{equation}

The single-point ITL score is the Gaussian mutual information between the candidate observation and the target effects:

\begin{equation}
s_{\mathrm{ITL}}(x;A)
=
-\frac{1}{2}
\log\!\left(
1-
\frac{
\Sigma_n(x,A)\Sigma_n(A,A)^{-1}\Sigma_n(A,x)
}{
\Sigma_n(x,x)+\tau^2
}
\right).
\end{equation}

The role of $\Sigma_n(A,A)^{-1}$ is to remove redundancy among correlated directions within $A$. Let $c_x=\Sigma_n(A,x)$ and $K_A=\Sigma_n(A,A)$. The ITL numerator is then $c_x^\top K_A^{-1}c_x$. If $K_A=Q\Lambda Q^\top$, this term equals $\|\Lambda^{-1/2}Q^\top c_x\|_2^2$, measuring the coupling strength between the candidate and $A$ along whitened target directions. Thus, exact ITL requires target-target covariance, whereas VTL only requires target-candidate cross-covariance and candidate marginal variance.

These formulas also specify the quantities required from the posterior interface. Single-point ITL/VTL scores are determined by target uncertainty, candidate marginal variance, observation noise, and accessible-target coupling. Within a batch, marginal scores are recomputed after repeatedly applying covariance-only conditioning. The full batch objective, conditional greedy procedure, and applicable conditions are given in Appendix~\ref{appb}.

This also explains why the posterior interface is central to the method: TALK does not require the surrogate to be Kermut, but it does require the selection posterior to provide sufficient target uncertainty and accessible-target coupling.

\section{Conditional Batch Construction}
\label{appb}

This appendix supplements the objective function, conditional greedy construction, and applicability conditions for fixed-budget batch selection. Protein variant experiments typically design, synthesize, and measure multiple variants in the same round; before that batch is completed, true within-batch labels cannot be used to revise later choices. Thus, a single-round wet-lab design corresponds to a non-adaptive batch problem: given the current observations $\mathcal{D}_n$, accessible pool $U_n$, target space $A$, and batch budget $B$, TALK outputs

\begin{table}[htbp]
\centering
\scriptsize
\resizebox{\linewidth}{!}{%
\begin{tabular}{lll}
\toprule
Component & Required posterior objects & Role in TALK \\
\midrule
Prediction on $A$ & $\mu_n(A)$, $\operatorname{diag}\Sigma_n(A,A)$ & Evaluate target predictions and uncertainty \\
Sample variance baseline & $\operatorname{diag}\Sigma_n(U_n,U_n)$ & Select points with the largest marginal uncertainty in the current query pool \\
VTL & $\Sigma_n(A,U_n)$, $\operatorname{diag}\Sigma_n(U_n,U_n)$, $\tau^2$ & Compute target-space trace variance reduction \\
Exact ITL & $\Sigma_n(A,A)$, $\Sigma_n(A,U_n)$, $\operatorname{diag}\Sigma_n(U_n,U_n)$, $\tau^2$ & Compute information gain between target effects and candidate observation \\
Dense conditional batch & $\Sigma_n(A,A)$, $\Sigma_n(A,U_n)$, $\Sigma_n(U_n,U_n)$, $\tau^2$ & Apply Schur-complement updates to the joint covariance over targets and candidates \\
Low-rank conditional batch & diagonal terms and low-rank factors for $A$ and $U_n$ & Approximately maintain covariance updates without explicitly storing large matrices \\
\bottomrule
\end{tabular}%
}
\end{table}

\begin{equation}
Q_n\subseteq U_n,\qquad |Q_n|=B.
\end{equation}

All elements in the batch share the same selection posterior; true labels are added to the next-round training set only after the batch measurement is complete. The goal of batch construction is to maximize the overall posterior improvement that the whole batch provides for $A$. In the main text, $G_n(Q;A)$ denotes the overall acquisition score of batch $Q$ for target space $A$. For ITL, the batch objective is

\begin{equation}
G_n^{\mathrm{ITL}}(Q;A)
=
I(f_A;y_Q\mid\mathcal{D}_n).
\end{equation}

For VTL, the batch objective is target-space trace variance reduction:

\begin{equation}
G_n^{\mathrm{VTL}}(Q;A)
=
\operatorname{tr}\Sigma_n(A,A)
-
\operatorname{tr}\operatorname{Var}(f_A\mid\mathcal{D}_n,y_Q).
\end{equation}

Both objectives define batch value on $A$: the quality of a batch depends on whether measuring these accessible variants improves posterior judgment over the target effects.

Solving

\begin{equation}
Q_n^\star
=
\arg\max_{Q\subseteq U_n,\ |Q|=B}
G_n(Q;A)
\end{equation}

directly would require enumerating many candidate batches. TALK uses conditional greedy construction as an executable approximation. Let

\begin{equation}
Q^{(j)}=\{x_1,\ldots,x_{j-1}\}
\end{equation}

denote the elements already selected into the current batch. At step $j$, TALK uses the covariance $\Sigma^{(j)}$ conditioned on $Q^{(j)}$ to recompute the marginal gain of all remaining candidates and selects

\begin{equation}
x_j
=
\arg\max_{x\in U_n\setminus Q^{(j)}}
s_g(x;A;\Sigma^{(j)}),
\end{equation}

where $g\in\{\mathrm{ITL},\mathrm{VTL}\}$. After selecting $x_j$, TALK updates $\Sigma^{(j)}$ using covariance-only Gaussian conditioning. Because the Gaussian posterior covariance update does not depend on the realized observation value, this step does not require and does not use unknown within-batch labels.

Concretely, for any finite sets $C,C'\subseteq X$, after selecting $x_j$ the conditional covariance is

\begin{equation}
\Sigma^{(j+1)}(C,C')
=
\Sigma^{(j)}(C,C')
-
\frac{
\Sigma^{(j)}(C,x_j)\Sigma^{(j)}(x_j,C')
}{
\Sigma^{(j)}(x_j,x_j)+\tau^2
}.
\end{equation}

This update does not access $y_{x_j}$, and is therefore applicable to one-shot batch planning. Compared with ranking once and taking top-$B$, conditional greedy updates the posterior covariance after each selection, so subsequent candidates must still have marginal value under the remaining target uncertainty.

Results in transductive active learning and submodular selection show that, if the normalized batch objective is nonnegative, monotone, and submodular, greedy selection inherits classical approximation guarantees for submodular maximization \citep{hubotter2024transductive,chen2013near,nemhauser1978analysis}. When features are correlated, the posterior is approximate, or the objective is not exactly submodular, weak submodularity provides a more permissive framework for interpreting greedy behavior \citep{das2018approximate}. These conditions give conditional greedy a clear optimization interpretation beyond purely empirical ranking rules.

For real protein posteriors, Sparse/FITC-style approximations, and complex $S/A$ splits, we do not assume that every assay strictly satisfies these conditions. TALK therefore uses conditional greedy as an auditable non-redundant batch construction, and evaluates its empirical effect through the no-conditional comparison in Section~\ref{44}, the budget curves in Figure~\ref{fig1:pgmulti51-protocols}, and protocol-matched static references.

\section{Scalable Target-Posterior Implementation}
\label{appc}

This appendix describes how TALK constructs the selection posterior under different target-size regimes, focusing on implementation paths and applicable scenarios. Detailed complexity analysis is given in Appendix~\ref{appe}. The source of Kermut posterior quantities is described in Appendix~\ref{appd}. Sparse/FITC-style approximation is used only as a scalable implementation path for large target panels.

\subsection{Target-size Regimes}

The target space $A$ in TALK may be a small finite panel, a large ProteinGym split, or a combinatorial mutation library. Different scales require different posterior representations:

\begin{itemize}
\item \textbf{Small finite $A$}: one can directly construct the dense covariance on $A\cup U_n$, supporting exact ITL, VTL, and dense conditional batch updates.
\item \textbf{Large finite $A$}: exact target-candidate cross-covariance is often still needed, but the full $\Sigma(A,A)$ or $\Sigma(U_n,U_n)$ may be too large, requiring chunking or low-rank representations.
\item \textbf{Very large $A$ or $U_n$}: when dense target-candidate covariance is also difficult to store explicitly, inducing-point or low-rank representations are needed to approximately maintain coupling and marginal uncertainty.
\end{itemize}

The core requirement is that the selection posterior must at least provide uncertainty on $A$ and posterior coupling between accessible candidates and $A$.

In practice, backend choice is mainly determined by target size, candidate size, whether exact ITL is required, and available memory. The batch budget $B$ mainly affects the number of conditioning updates, and is usually not the primary factor in choosing the backend.

\subsection{Dense Exact Posterior}

When $|A|+|U_n|$ is manageable, the most direct implementation constructs a dense Gaussian posterior on $A\cup U_n$. This representation provides

\begin{equation}
\Sigma(A,A),\qquad
\Sigma(A,U_n),\qquad
\Sigma(U_n,U_n),
\end{equation}

as well as $\mu_A,\mu_U$ and marginal variances. Dense representation is conceptually the simplest: exact ITL can use the full $\Sigma(A,A)$, VTL can use the full cross-covariance, and conditional batch updates can be applied to the joint covariance by Schur-complement updates.

The limitation of the dense exact posterior is that memory grows quadratically with $|A|+|U_n|$. It is therefore suited to small target panels, small-scale exact validation, and confirmatory runs that require the full covariance.

\subsection{Exact Chunked Cross Posterior}

When the target panel is large but an exact GP posterior is still desired, chunked construction can be used. This path computes the joint posterior between target chunks and the candidate pool block by block, yielding

\begin{equation}
\mu_A,\ \mu_U,\ \operatorname{diag}\Sigma(A,A),\ \operatorname{diag}\Sigma(U_n,U_n),\ \Sigma(A,U_n).
\end{equation}

If memory permits, $\Sigma(A,A)$ can also be constructed in chunks. Provided that $U_n$ itself can still be represented densely, this allows exact ITL and dense batch updates for medium-sized targets. If $\Sigma(A,A)$ is not constructed, the backend is better suited to VTL or other scores that only require cross-covariance.

The role of the chunked posterior is to split a large joint covariance query into manageable blocks. It does not change the mathematical definition of the GP posterior; it only changes how the quantities are computed. Thus, when resources permit, the chunked exact posterior is an implementation extension of the dense exact posterior.

\subsection{Sparse/FITC-style Low-rank Backend}

For very large target panels, full dense covariance is often infeasible. In this paper, Sparse/FITC-style refers to an inducing-point sparse GP-style posterior approximation: a small number of inducing variables and a diagonal correction are used to approximate the posterior covariance on finite sets. This backend stores covariance in the form

\begin{equation}
\Sigma(C,C)\approx D_C+L_C L_C^\top
\end{equation}

where $D_C=\operatorname{diag}(d_C)$ is a diagonal correction and $L_C$ is a low-rank factor. For $A$ and $U_n$, the implementation maintains

\begin{equation}
D_A,\ D_U,\ L_A,\ L_U,
\end{equation}

which implicitly provide target uncertainty, candidate uncertainty, and target-candidate coupling. For target-candidate cross-covariance, the diagonal correction does not contribute off-diagonal coupling, so acquisition uses the low-rank term

\begin{equation}
\Sigma(A,U_n)\approx L_A L_U^\top.
\end{equation}

Marginal variances are still given by the diagonal of $D_C+L_C L_C^\top$, so the candidate denominator and target uncertainty include the diagonal correction. The inducing set controls the expressive capacity of the low-rank posterior. In practice, inducing candidates come from a training-side available fitting pool that excludes target labels, or from other selection-time available candidates, and representative points are selected by embedding-space coverage. In cold-start settings, they can be constructed from protocol-allowed measurable candidates or non-target label information. Candidate coverage caps and chunk size jointly control coverage and memory pressure for large target panels. In retrospective evaluation, held-out target labels do not enter fitting or inducing selection.

\subsection{Stochastic Target-Space Approximation}

In protein design settings, the target space $A$ may be much larger than the target panels used in our main retrospective benchmarks. Examples include all higher-order combinations over a set of mutable sites, a computationally generated design library, or a large panel of variants passing structural or functional filters. In these settings, even chunked construction of $\Sigma(A,U_n)$ can be too expensive. A complementary approximation is to replace full enumeration of $A$ with a sampled target subset $\tilde A_K \subset A$ when computing the selection score.

Let $P_A$ be a distribution over variants in the target space, constructed from information available before selection. At a selection step, sample $K$ target variants from $P_A$ and compute the ITL or VTL score using $\tilde A_K$ in place of $A$. This gives a Monte Carlo approximation to the target-space objective while preserving the TALK constraint that labels from the target space are not used for selection.

The choice of $P_A$ should reflect how the target space is meant to be used. A uniform distribution treats all target variants equally. A stratified distribution can preserve coverage across mutation orders, sites, or design constraints. A protein-language-model-guided distribution can bias sampling toward variants that are plausible under a protein language model, diverse in embedding space, or promising under a zero-shot score, provided that these quantities are computed without target labels. Such PLM-guided sampling does not replace the posterior coupling used by TALK; it only chooses which parts of an ultra-large target space are represented in the stochastic approximation.

The reliability of this approximation depends on whether $P_A$ covers the target regions whose prediction quality matters. If important regions receive little probability mass, the sampled target subset can underrepresent the uncertainty that TALK should reduce. Thus, stochastic target-space sampling should report the sampling rule, sample size $K$, and any PLM score or embedding used to construct $P_A$. This extension is orthogonal to the posterior backend: the sampled subset can be used with dense, chunked, or sparse posterior representations. In this paper, the main experiments use finite target spaces; stochastic target-space sampling is a scalable extension for ultra-large combinatorial target spaces.

\section{Kermut Posterior Interface}
\label{appd}

This appendix describes how the Kermut posterior provides the predictive distribution quantities required by TALK \citep{groth2024kermut}. Kernel validity, benchmark details, and full model ablations for Kermut are not repeated here. The TALK/Kermut pipeline fits a posterior using variant sequences, sequence embeddings, zero-shot scores, structure-kernel inputs, observed labels, and a Gaussian likelihood noise estimate. Together, these objects determine the posterior mean, marginal variance, and covariance. For TALK, the role of the Kermut posterior is to provide posterior coupling between accessible variants and target variants.

Kermut's mean function can be a constant mean or a zero-shot mean, for example

\begin{equation}
m(x)=\alpha f_0(x)+\beta,
\end{equation}

where $f_0(x)$ is a zero-shot fitness estimate. Its covariance function is provided jointly by sequence and structure kernels. The model uses a Gaussian likelihood and learns kernel hyperparameters and noise through marginal likelihood.

TALK reads a finite set of predictive distribution quantities from the Kermut posterior. For target variants, it uses posterior mean and marginal variance to report prediction performance and uncertainty. For unqueried accessible candidates, it uses candidate marginal variance and observation-noise variance to quantify the noise scale of a measurement. For acquisition scores, the key quantity is posterior covariance between target variants and accessible candidates. Exact ITL also requires posterior covariance within the target panel to remove redundancy among target directions, whereas VTL mainly depends on target-candidate covariance and candidate marginal variance. For large target panels, Kermut can approximate these quantities with a Sparse/FITC-style representation, using diagonal terms and low-rank factors to represent target uncertainty, candidate uncertainty, and their coupling. Implementation paths for different scales are described in Appendix~\ref{appc}.

Prediction and acquisition use overlapping but not identical parts of the posterior. Prediction mainly uses $\mu_n(A)$ and marginal uncertainty; ITL/VTL mainly use posterior covariance, accessible-target coupling, and observation-noise variance. A higher predictive mean is not itself a higher measurement value, which is a main distinction between top-mean/UCB and ITL/VTL.

Observed labels are standardized within each assay using the current training labels. The Gaussian likelihood provides a learned homoscedastic noise estimate $\tau^2$, which enters the acquisition denominator

\begin{equation}
\Sigma_n(x,x)+\tau^2,
\end{equation}

the predictive variance of the noisy observation $y_x$. Larger observation noise reduces the expected information provided by a candidate measurement.

The reliability of the Kermut posterior depends on consistent alignment between protein inputs and the variant table. Zero-shot scores, sequence embeddings, structure features, observed labels, and train/selection/evaluation splits must all refer to the same canonical variant table. Otherwise, posterior coupling cannot be interpreted as correlation among the same set of protein variants.

We use posterior-quality information only to the extent needed for selection: marginal variances of targets and candidates, the strength and concentration of target-candidate coupling, and whether target variance reduction saturates with budget. A more detailed support interpretation is given in Appendix~\ref{appf}.

The selection posterior can be provided by Bayesian surrogates other than Kermut. If a last-layer model, NTK approximation, linearized embedding model, or another Bayesian surrogate can provide posterior covariance, marginal variance, and observation-noise estimates between target variants and accessible candidates, the same ITL/VTL scores can be computed. We use Kermut as the main instantiation because it is designed for protein variant effect modeling and directly provides sequence/structure-aware covariance. Alternative posterior substrates are treated as interface-level extensions, not as the main experimental focus of this paper.

\section{Computational Complexity and Memory Scaling}
\label{appe}

This appendix summarizes the main time and memory scales of different posterior backends. Let $m=|A|$ be the target size, $u=|U_n|$ the number of current candidates, $b$ the batch budget, $r$ the low-rank factor rank, and $c$ the chunk size.

\begin{table}[htbp]
\centering
\scriptsize
\resizebox{\linewidth}{!}{%
\begin{tabular}{lrrrl}
\toprule
Backend & Main storage & Main cost per score & Batch update & Use case \\
\midrule
Dense exact & $O((m+u)^2)$ & ITL depends on solves with $\Sigma(A,A)$; VTL is about $O(mu)$ & dense Schur update; quadratic in joint covariance size & small panels, exact debugging \\
Exact chunked cross & $O(cu)$, or $O(mu)$ if storing $\Sigma(A,U)$ & chunked cross-covariance computation; VTL can be accumulated in streaming form & dense update only if joint covariance is constructed & medium targets, exact cross-covariance \\
Sparse/FITC-style low-rank & $O((m+u)r)$ & cross-coupling about $O(mur)$ or chunked factor products & low-rank transform, mainly depends on $r$ and $b$ & large target panels \\
\bottomrule
\end{tabular}%
}
\end{table}

Dense exact inference has the most direct mathematical form, but memory becomes prohibitive as the target panel grows. Chunked exact inference preserves exact posterior queries while controlling peak memory through blocks. The Sparse/FITC-style backend uses an inducing representation to reduce storage pressure for target-candidate coupling, at the cost that acquisition scores depend on inducing-set coverage and approximation quality \citep{snelson2005sparse,titsias2009variational}.

Complexity reporting supports experimental interpretation. When exact ITL is infeasible, it is important to state that the bottleneck comes from storing or solving with $\Sigma(A,A)$, not from an incomplete acquisition objective. When using a Sparse/FITC-style backend, inducing count, target cap, and chunk size provide necessary context for interpreting the approximation path.

\section{Protein-space Support Diagnostics and Failure Modes}
\label{appf}

This appendix explains TALK's gains and boundaries from the perspective of protein-space support. The central question is whether the accessible pool allowed by the current protocol can, through posterior coupling, support the main uncertainty on the target set $A$. When such support exists, ITL/VTL can prioritize target-relevant uncertainty reduction under a limited budget. When support is weak, a better acquisition rule cannot obtain information from $S$ that is not present there.

A direct diagnostic is the residual target uncertainty after ``fully measuring the accessible pool.'' Given the current posterior and the currently queryable candidate set $U_n$, if all variants in $U_n$ were observed under the same Gaussian noise model, the conditional covariance on $A$ would be

\begin{equation}
\Sigma_n^{A\mid U}
=
\Sigma_n(A,A)
-
\Sigma_n(A,U_n)
\left(\Sigma_n(U_n,U_n)+\tau^2 I\right)^{-1}
\Sigma_n(U_n,A).
\end{equation}

When $\operatorname{tr}\Sigma_n(A,A)>0$, we normalize the reducible target
variance as
\begin{equation}
R =
\frac{\operatorname{tr}\Sigma_n(A,A)-\operatorname{tr}\Sigma_n(A, U)}
     {\operatorname{tr}\Sigma_n(A,A)}.
\end{equation}
Under the declared posterior, $R$ summarizes the fraction of target
variance reducible by observing the whole accessible pool. It is not a
finite-budget performance guarantee or a universal cross-assay success
threshold. It may be used to compare candidate pools or target panels under
the same model, but it does not validate a misspecified posterior.

Here $\operatorname{tr}\Sigma_n(A,A)-\operatorname{tr}\Sigma_n^{A\mid U}$ is the maximum amount of target uncertainty that the current accessible pool can explain under the posterior, while $\operatorname{tr}\Sigma_n^{A\mid U}$ is the residual target uncertainty that remains even after fully measuring the accessible pool. TALK aims to approach the reducible portion with a budget smaller than the full accessible pool. It cannot eliminate uncertainty left by the target space itself, the expressivity of the posterior model, or assay-specific residual structure.

The static reference is the performance-side counterpart of this diagnostic in retrospective experiments: it fits the model using the complete measurable train-side labels allowed by the protocol and evaluates on the same target space. If an active curve quickly approaches the static reference, the main limitation is likely accessible-data support rather than batch construction. If a clear gap remains between the active curve and the static reference, possible causes include insufficient budget, batch redundancy, or acquisition scores that fail to cover target-relevant posterior directions. The static reference is an empirical reference line, not a mathematically unbreakable upper bound. In small accessible pools or high-noise tasks, finite-subset curves may locally exceed it, but the overall gap remains useful for distinguishing selection problems from support problems.

In protein variant spaces, support failure typically has three sources. First, posterior coupling between accessible candidates and target variants may be weak; for example, low-order mutations may fail to explain specific site interactions in some higher-order combinations. Second, coupling may be concentrated in only a few target directions; after a batch covers those directions, remaining target uncertainty decreases slowly. Third, the kernel, embedding, or noise model may fail to express key structure in the assay, causing posterior covariance to underestimate or misplace true mutation-space relatedness. These cases lead active curves to saturate early. TALK may still outperform random or local-uncertainty baselines, but its gains are jointly limited by the measurable space and the posterior model.

\section{Budget Coverage and Random-Equivalent Sample Efficiency}
\label{appg}

Section~\ref{42} reports active selection curves using absolute measurement budgets. Because the available sampling spaces differ across assays in ProteinGym-MM, the same budget corresponds to different wet-lab sampling intensities across tasks. We define budget share as

\begin{equation}
\mathrm{budget\ share}(B)=\frac{B}{|U_{\mathrm{available}}|},
\end{equation}

where $U_{\mathrm{available}}$ is the unlabeled sampling pool that can be actively selected under the protocol. For canonical $1M\to2M$, $|U_{\mathrm{available}}|=|1M|$. For the mixed-order pool protocol, $|U_{\mathrm{available}}|=|1M_{\mathrm{train}}\cup2M_{\mathrm{train}}|$. For the Singles-premeasured protocol, $|U_{\mathrm{available}}|=|2M_{\mathrm{train}}|$. In fold-aware settings, the statistical unit is an assay-fold task. If a high-budget checkpoint exceeds the sampleable double-mutant pool for some folds, that checkpoint is computed only over tasks where the budget is feasible.

\begin{figure}[htbp]
\centering
\includegraphics[width=\linewidth]{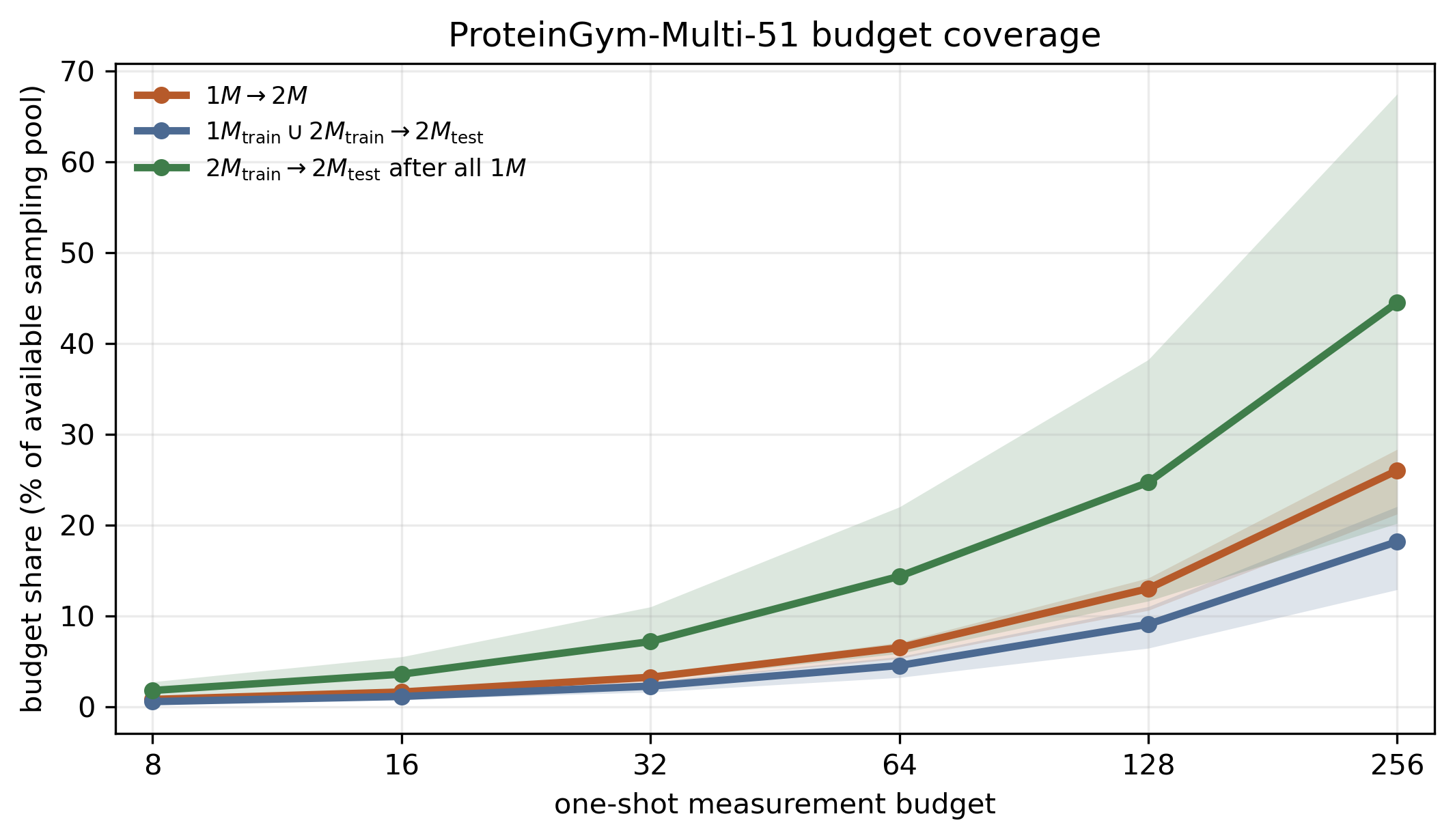}
\caption{ProteinGym-MM measurement budgets as fractions of assay-specific sampling pools. Lines indicate mean budget share across tasks, and shaded regions show the 25th--75th percentile range. The same absolute budget occupies a larger fraction of the double-mutant pool in the Singles-premeasured protocol, so comparisons of saturation speed across protocols should account for available pool size.}
\label{fig:budget-coverage}
\end{figure}

\begin{table}[htbp]
\centering
\scriptsize
\caption{Budget share summary for ProteinGym-MM protocols. Values are percentages of the protocol-specific available sampling pool. \texttt{n} is the number of tasks feasible at that budget. When an assay-fold has $|U_{\mathrm{available}}|<B$, that task is excluded from the coverage statistic at that budget and is not forced to sample more variants than the available pool in the corresponding active run.}
\label{tab:04-table-g-1-budget-share-summary-for-proteingym-mm}
\resizebox{\linewidth}{!}{%
\begin{tabular}{lrrrrrrrr}
\toprule
Setting & B & n & mean \% & min \% & p25 \% & median \% & p75 \% & max \% \\
\midrule
$1M\to2M$ & 8 & 51 & 0.81 & 0.59 & 0.66 & 0.77 & 0.89 & 1.29 \\
$1M\to2M$ & 16 & 51 & 1.63 & 1.18 & 1.33 & 1.55 & 1.77 & 2.58 \\
$1M\to2M$ & 32 & 51 & 3.25 & 2.36 & 2.65 & 3.09 & 3.54 & 5.15 \\
$1M\to2M$ & 64 & 51 & 6.50 & 4.72 & 5.30 & 6.18 & 7.09 & 10.31 \\
$1M\to2M$ & 128 & 51 & 13.01 & 9.43 & 10.61 & 12.37 & 14.17 & 20.61 \\
$1M\to2M$ & 256 & 51 & 26.01 & 18.87 & 21.22 & 24.73 & 28.34 & 41.22 \\
mixed-order pool & 8 & 255 & 0.57 & 0.14 & 0.40 & 0.53 & 0.69 & 1.25 \\
mixed-order pool & 16 & 255 & 1.14 & 0.29 & 0.81 & 1.05 & 1.38 & 2.50 \\
mixed-order pool & 32 & 255 & 2.27 & 0.57 & 1.61 & 2.10 & 2.76 & 5.00 \\
mixed-order pool & 64 & 255 & 4.54 & 1.15 & 3.22 & 4.20 & 5.51 & 10.00 \\
mixed-order pool & 128 & 255 & 9.09 & 2.29 & 6.45 & 8.41 & 11.02 & 20.00 \\
mixed-order pool & 256 & 255 & 18.17 & 4.58 & 12.90 & 16.82 & 22.05 & 40.00 \\
Singles-premeasured protocol & 8 & 255 & 1.80 & 0.18 & 0.73 & 1.40 & 2.75 & 9.09 \\
Singles-premeasured protocol & 16 & 255 & 3.59 & 0.35 & 1.46 & 2.80 & 5.50 & 18.18 \\
Singles-premeasured protocol & 32 & 255 & 7.18 & 0.70 & 2.92 & 5.59 & 11.00 & 36.36 \\
Singles-premeasured protocol & 64 & 255 & 14.36 & 1.40 & 5.85 & 11.19 & 21.99 & 72.73 \\
Singles-premeasured protocol & 128 & 245 & 24.71 & 2.80 & 11.64 & 21.96 & 38.21 & 74.42 \\
Singles-premeasured protocol & 256 & 230 & 44.50 & 5.61 & 20.19 & 36.57 & 67.46 & 96.97 \\
\bottomrule
\end{tabular}%
}
\end{table}

The three faithful GB1 tasks are single-task complete-landscape splits, so budget share can be computed directly from the FLIP train-side pool size. In \texttt{1-vs-rest}, the measurable pool contains \texttt{28} variants, and budgets \texttt{4/8/16/24/28} cover approximately \texttt{14.29\%/28.57\%/57.14\%/85.71\%/100.00\%}. In \texttt{2-vs-rest}, the measurable pool contains \texttt{424} variants, and budgets \texttt{8/16/32/64/128/256} cover about \texttt{1.89\%} to \texttt{60.38\%}. In \texttt{3-vs-rest}, the measurable pool contains \texttt{2990} variants, and the same budgets cover only about \texttt{0.27\%} to \texttt{8.56\%}. Thus, the meaning of budget differs substantially across GB1 tasks: \texttt{1-vs-rest} is closer to near-complete sampling of a small pool, while \texttt{3-vs-rest} better reflects higher-order active design under a limited budget.

The FLIP \texttt{low-vs-high} split for GB1 is not used as a main faithful task. This split is defined retrospectively by a true fitness threshold and uses label information that would be unavailable before the wet-lab experiment to partition measurable and target domains. The train/test sides also mix multiple mutation orders, and WT is placed on the test side. Thus, \texttt{low-vs-high} is better interpreted as a label-defined distribution-shift or fine-tuning stress test than as an executable wet-lab active-design protocol.

To translate coverage into a more direct experimental-efficiency interpretation, we also compute Random-equivalent budget saving. For each setting and each Random budget checkpoint $B$, we first take Random's Spearman at that budget. We then linearly interpolate on the monotone envelope of \texttt{max(ITL,VTL)} to estimate the minimum target-aware budget $B_{\mathrm{TA}}$ needed to reach the same Spearman. The sample saving rate is defined as

\begin{equation}
\mathrm{saving}(B)=1-\frac{B_{\mathrm{TA}}}{B}.
\end{equation}

This quantity is a post hoc curve-level sample-efficiency summary and does not introduce a new training protocol. On ProteinGym-MM, the best of ITL/VTL saves on average \texttt{78.5\%}, \texttt{61.1\%}, and \texttt{34.0\%} of measurements needed to match Random's Spearman in the canonical $1M\to2M$, mixed-order pool, and Singles-premeasured protocols, respectively. The savings decrease from cold-start to warm-start settings, consistent with the intuition that richer initial information leaves less room for active design to improve. On GB1, the average savings for \texttt{2-vs-rest} and \texttt{3-vs-rest} are \texttt{60.0\%} and \texttt{59.2\%}, showing that target-aware selection also reduces the number of measurements needed to reach Random-equivalent ranking performance in complete-landscape tasks. The \texttt{1-vs-rest} pool contains only \texttt{28} variants, and early fluctuations from a few points can temporarily favor Random, so it is not well summarized by the mean over all budgets. At the final checkpoint, the best of ITL/VTL needs about \texttt{22.0} samples to reach Random's Spearman at \texttt{B=28}, corresponding to \texttt{21.5\%} sample saving.

\begin{table}[htbp]
\centering
\scriptsize
\caption{Random-equivalent sample savings for best ITL/VTL. \texttt{mean\ saving} and \texttt{median\ saving} are computed over nonzero budget checkpoints in each setting; \texttt{final\ saving} is the budget reduction of best ITL/VTL needed to reach final-budget Random Spearman.}
\label{tab:05-table-g-2-random-equivalent-sample-savings-for-b}
\resizebox{\linewidth}{!}{%
\begin{tabular}{lrrrrrr}
\toprule
Setting & n budgets & mean saving \% & median saving \% & final Random B & best ITL/VTL B for final Random & final saving \% \\
\midrule
ProteinGym-MM $1M\to2M$ & 6 & 78.5 & 75.5 & 256 & 53.0 & 79.3 \\
ProteinGym-MM mixed-order pool protocol & 6 & 61.1 & 61.2 & 256 & 91.6 & 64.2 \\
ProteinGym-MM Singles-premeasured protocol & 6 & 34.0 & 36.8 & 256 & 192.0 & 25.0 \\
GB1 \texttt{1-vs-rest} & 5 & -31.3 & 21.5 & 28 & 22.0 & 21.5 \\
GB1 \texttt{2-vs-rest} & 6 & 60.0 & 64.3 & 256 & 118.7 & 53.6 \\
GB1 \texttt{3-vs-rest} & 6 & 59.2 & 65.9 & 256 & 110.7 & 56.8 \\
\bottomrule
\end{tabular}%
}
\end{table}

\section{Full Budget-Performance Matrices for Sections 4.2 and 4.3}
\label{apph}

This appendix summarizes the complete method-by-budget performance matrices for the main experiments in Sections 4.2 and 4.3. Tables H.1--H.6 report macro-average Spearman correlation on the corresponding target space. Tables H.7--H.12 report top-5\% target recovery, defined as the overlap between predicted top-5\% target variants and true top-5\% target variants. ProteinGym-MM results are averaged over assays or assay-fold tasks, whereas GB1 results are single-task metrics. For the Random baseline, five random seeds are first averaged within the same task-budget pair before entering the macro-average. ITL, VTL, SampleVariance, top-mean, and UCB are deterministic ranking rules, with one ranking result per task-budget. Table order follows the main-text narrative.

\begin{table}[htbp]
\centering
\scriptsize
\caption{ProteinGym-MM canonical $1M\to2M$: Spearman by method and budget.}
\label{tab:06-table-h-1-proteingym-mm-canonical-1m-to2m-spearm}
\resizebox{\linewidth}{!}{%
\begin{tabular}{lrrrrrrr}
\toprule
Method & $B=0$ & $B=8$ & $B=16$ & $B=32$ & $B=64$ & $B=128$ & $B=256$ \\
\midrule
ITL & 0.1888 & 0.3203 & 0.4319 & 0.5190 & 0.5931 & 0.6428 & 0.6518 \\
VTL & 0.1888 & 0.3308 & 0.4310 & 0.5156 & 0.5756 & 0.6172 & 0.6432 \\
SampleVariance & 0.1888 & 0.2579 & 0.3537 & 0.4441 & 0.4931 & 0.5321 & 0.5812 \\
Random & 0.1888 & 0.1887 & 0.2690 & 0.3557 & 0.4328 & 0.5103 & 0.5675 \\
-Random SD (n=5) & 0.0000 & 0.0257 & 0.0218 & 0.0089 & 0.0243 & 0.0141 & 0.0065 \\
top-mean & 0.1888 & 0.1152 & 0.1454 & 0.2115 & 0.2540 & 0.3297 & 0.3920 \\
UCB & 0.1888 & 0.1261 & 0.1463 & 0.2101 & 0.2457 & 0.3320 & 0.3956 \\
\bottomrule
\end{tabular}%
}
\end{table}

\begin{table}[htbp]
\centering
\scriptsize
\caption{ProteinGym-MM mixed-order pool protocol: Spearman by method and budget.}
\label{tab:07-table-h-2-proteingym-mm-mixed-order-pool-protoco}
\resizebox{\linewidth}{!}{%
\begin{tabular}{lrrrrrrr}
\toprule
Method & $B=0$ & $B=8$ & $B=16$ & $B=32$ & $B=64$ & $B=128$ & $B=256$ \\
\midrule
ITL & 0.1860 & 0.4065 & 0.5649 & 0.6800 & 0.7696 & 0.8302 & 0.8702 \\
VTL & 0.1860 & 0.3922 & 0.5540 & 0.6928 & 0.7855 & 0.8417 & 0.8778 \\
SampleVariance & 0.1860 & 0.3830 & 0.5060 & 0.5985 & 0.6485 & 0.6823 & 0.7163 \\
Random & 0.1860 & 0.2599 & 0.3810 & 0.5157 & 0.6389 & 0.7413 & 0.8098 \\
-Random SD (n=5) & 0.0000 & 0.0195 & 0.0115 & 0.0075 & 0.0053 & 0.0057 & 0.0035 \\
UCB & 0.1860 & 0.1230 & 0.1333 & 0.1961 & 0.2731 & 0.3544 & 0.4604 \\
top-mean & 0.1860 & 0.1255 & 0.1290 & 0.1965 & 0.2632 & 0.3517 & 0.4579 \\
\bottomrule
\end{tabular}%
}
\end{table}

\begin{table}[htbp]
\centering
\scriptsize
\caption{ProteinGym-MM Singles-premeasured protocol: Spearman by method and budget.}
\label{tab:08-table-h-3-proteingym-mm-singles-premeasured-prot}
\resizebox{\linewidth}{!}{%
\begin{tabular}{lrrrrrrr}
\toprule
Method & $B=0$ & $B=8$ & $B=16$ & $B=32$ & $B=64$ & $B=128$ & $B=256$ \\
\midrule
ITL & 0.6371 & 0.7034 & 0.7387 & 0.7853 & 0.8175 & 0.8670 & 0.8907 \\
VTL & 0.6371 & 0.7095 & 0.7498 & 0.7887 & 0.8255 & 0.8635 & 0.8838 \\
SampleVariance & 0.6371 & 0.6867 & 0.7167 & 0.7509 & 0.7935 & 0.8453 & 0.8831 \\
Random & 0.6371 & 0.6836 & 0.7200 & 0.7575 & 0.7971 & 0.8460 & 0.8801 \\
-Random SD (n=5) & 0.0000 & 0.0039 & 0.0019 & 0.0026 & 0.0017 & 0.0014 & 0.0011 \\
UCB & 0.6371 & 0.6047 & 0.6210 & 0.6301 & 0.6583 & 0.6915 & 0.7288 \\
top-mean & 0.6371 & 0.6050 & 0.6127 & 0.6281 & 0.6567 & 0.6859 & 0.7249 \\
\bottomrule
\end{tabular}%
}
\end{table}

\begin{table}[htbp]
\centering
\scriptsize
\caption{GB1 \texttt{1-vs-rest}: Spearman by method and budget.}
\label{tab:09-table-h-4-gb1-1-vs-rest-spearman-by-method-and-b}
\resizebox{\linewidth}{!}{%
\begin{tabular}{lrrrrr}
\toprule
Method & $B=4$ & $B=8$ & $B=16$ & $B=24$ & $B=28$ \\
\midrule
ITL & 0.2183 & 0.0939 & 0.2230 & 0.3828 & 0.3839 \\
VTL & 0.1351 & 0.1021 & 0.3702 & 0.3886 & 0.3839 \\
SampleVariance & 0.1042 & 0.0475 & 0.2571 & 0.3740 & 0.3839 \\
Random mean & 0.1608 & 0.1700 & 0.2998 & 0.3656 & 0.3839 \\
-Random SD (n=5) & 0.1406 & 0.1491 & 0.0855 & 0.0601 & 0.0000 \\

top-mean & 0.3235 & 0.3207 & 0.4143 & 0.4501 & 0.3839 \\
UCB & 0.3235 & 0.3207 & 0.4143 & 0.4501 & 0.3839 \\
\bottomrule
\end{tabular}%
}
\end{table}

\begin{table}[htbp]
\centering
\scriptsize
\caption{GB1 \texttt{2-vs-rest}: Spearman by method and budget.}
\label{tab:10-table-h-5-gb1-2-vs-rest-spearman-by-method-and-b}
\resizebox{\linewidth}{!}{%
\begin{tabular}{lrrrrrr}
\toprule
Method & $B=8$ & $B=16$ & $B=32$ & $B=64$ & $B=128$ & $B=256$ \\
\midrule
ITL & 0.3743 & 0.2138 & 0.1817 & 0.0188 & 0.4245 & 0.4879 \\
VTL & 0.3890 & 0.2327 & 0.3042 & 0.2632 & 0.4852 & 0.5851 \\
SampleVariance & 0.3753 & 0.2183 & 0.1817 & 0.1678 & 0.4664 & 0.4868 \\
Random & -0.1449 & -0.0820 & -0.0206 & 0.0930 & 0.3426 & 0.4712 \\
-Random SD (n=5) & 0.0765 & 0.2444 & 0.1410 & 0.1435 & 0.0334 & 0.0427 \\
top-mean & 0.2418 & -0.0885 & 0.2643 & 0.3047 & -0.0681 & 0.5113 \\
UCB & 0.2418 & 0.2900 & 0.2636 & 0.2836 & -0.0102 & 0.5095 \\
\bottomrule
\end{tabular}%
}
\end{table}

\begin{table}[htbp]
\centering
\scriptsize
\caption{GB1 \texttt{3-vs-rest}: Spearman by method and budget.}
\label{tab:11-table-h-6-gb1-3-vs-rest-spearman-by-method-and-b}
\resizebox{\linewidth}{!}{%
\begin{tabular}{lrrrrrr}
\toprule
Method & $B=8$ & $B=16$ & $B=32$ & $B=64$ & $B=128$ & $B=256$ \\
\midrule
ITL & 0.4993 & 0.5561 & 0.4622 & 0.5764 & 0.6193 & 0.6973 \\
VTL & 0.4006 & 0.6187 & 0.6520 & 0.6392 & 0.6905 & 0.7448 \\
SampleVariance & 0.5277 & 0.5359 & 0.4587 & 0.5711 & 0.6232 & 0.6897 \\
Random & 0.2241 & 0.0993 & 0.4111 & 0.5177 & 0.6066 & 0.6801 \\
-Random SD (n=5) & 0.3653 & 0.3971 & 0.1978 & 0.0728 & 0.0403 & 0.0440 \\
top-mean & 0.4385 & 0.5516 & 0.4938 & 0.5038 & 0.4586 & 0.6046 \\
UCB & 0.4385 & 0.5516 & 0.4938 & 0.5038 & 0.4586 & 0.6046 \\
\bottomrule
\end{tabular}%
}
\end{table}

\begin{table}[htbp]
\centering
\scriptsize
\caption{ProteinGym-MM canonical $1M\to2M$: top-5\% target recovery by method and budget.}
\label{tab:12-table-h-7-proteingym-mm-canonical-1m-to2m-top-5-}
\resizebox{\linewidth}{!}{%
\begin{tabular}{lrrrrrrr}
\toprule
Method & $B=0$ & $B=8$ & $B=16$ & $B=32$ & $B=64$ & $B=128$ & $B=256$ \\
\midrule
ITL & 0.1676 & 0.2707 & 0.3204 & 0.3719 & 0.4243 & 0.4686 & 0.4563 \\
VTL & 0.1676 & 0.1784 & 0.2642 & 0.3445 & 0.3949 & 0.4408 & 0.4665 \\
SampleVariance & 0.1676 & 0.2099 & 0.2392 & 0.2888 & 0.3199 & 0.3444 & 0.4000 \\
Random & 0.1676 & 0.1614 & 0.1988 & 0.2373 & 0.2751 & 0.3196 & 0.3668 \\
-Random SD (n=5) & 0.0000 & 0.0201 & 0.0047 & 0.0114 & 0.0116 & 0.0172 & 0.0132 \\
UCB & 0.1676 & 0.1470 & 0.1496 & 0.1595 & 0.2216 & 0.2485 & 0.2780 \\
top-mean & 0.1676 & 0.1350 & 0.1497 & 0.1546 & 0.2253 & 0.2474 & 0.2790 \\
\bottomrule
\end{tabular}%
}
\end{table}

\begin{table}[htbp]
\centering
\scriptsize
\caption{ProteinGym-MM mixed-order pool protocol: top-5\% target recovery by method and budget.}
\label{tab:13-table-h-8-proteingym-mm-mixed-order-pool-protoco}
\resizebox{\linewidth}{!}{%
\begin{tabular}{lrrrrrrr}
\toprule
Method & $B=0$ & $B=8$ & $B=16$ & $B=32$ & $B=64$ & $B=128$ & $B=256$ \\
\midrule
ITL & 0.1697 & 0.2692 & 0.3598 & 0.4564 & 0.5461 & 0.6057 & 0.6590 \\
VTL & 0.1697 & 0.1996 & 0.2817 & 0.4212 & 0.5221 & 0.6084 & 0.6551 \\
SampleVariance & 0.1697 & 0.2484 & 0.3470 & 0.3893 & 0.4329 & 0.4611 & 0.4875 \\
Random & 0.1697 & 0.1809 & 0.2318 & 0.3099 & 0.3949 & 0.4730 & 0.5590 \\
-Random SD (n=5) & 0.0000 & 0.0143 & 0.0063 & 0.0093 & 0.0090 & 0.0051 & 0.0111 \\
UCB & 0.1697 & 0.1287 & 0.1322 & 0.1664 & 0.2250 & 0.2599 & 0.3177 \\
top-mean & 0.1697 & 0.1294 & 0.1377 & 0.1718 & 0.2175 & 0.2576 & 0.3136 \\
\bottomrule
\end{tabular}%
}
\end{table}

\begin{table}[htbp]
\centering
\scriptsize
\caption{ProteinGym-MM Singles-premeasured protocol: top-5\% target recovery by method and budget.}
\label{tab:14-table-h-9-proteingym-mm-singles-premeasured-prot}
\resizebox{\linewidth}{!}{%
\begin{tabular}{lrrrrrrr}
\toprule
Method & $B=0$ & $B=8$ & $B=16$ & $B=32$ & $B=64$ & $B=128$ & $B=256$ \\
\midrule
ITL & 0.4537 & 0.4728 & 0.5009 & 0.5357 & 0.5854 & 0.6431 & 0.6949 \\
VTL & 0.4537 & 0.4791 & 0.5028 & 0.5446 & 0.5983 & 0.6428 & 0.6928 \\
SampleVariance & 0.4537 & 0.4810 & 0.5026 & 0.5125 & 0.5564 & 0.6010 & 0.6619 \\
Random & 0.4537 & 0.4730 & 0.4913 & 0.5216 & 0.5636 & 0.6124 & 0.6644 \\
-Random SD (n=5) & 0.0000 & 0.0028 & 0.0071 & 0.0114 & 0.0080 & 0.0061 & 0.0035 \\
top-mean & 0.4537 & 0.4049 & 0.4184 & 0.4503 & 0.4873 & 0.5032 & 0.5327 \\
UCB & 0.4537 & 0.4024 & 0.4184 & 0.4557 & 0.4938 & 0.5125 & 0.5417 \\
\bottomrule
\end{tabular}%
}
\end{table}

\begin{table}[htbp]
\centering
\scriptsize
\caption{GB1 \texttt{1-vs-rest}: top-5\% target recovery by method and budget.}
\label{tab:15-table-h-10-gb1-1-vs-rest-top-5-target-recovery-b}
\resizebox{\linewidth}{!}{%
\begin{tabular}{lrrrrr}
\toprule
Method & $B=4$ & $B=8$ & $B=16$ & $B=24$ & $B=28$ \\
\midrule
ITL & 0.0183 & 0.1193 & 0.1399 & 0.1583 & 0.1583 \\
VTL & 0.0688 & 0.0780 & 0.1261 & 0.1468 & 0.1583 \\
SampleVariance & 0.0069 & 0.1170 & 0.1514 & 0.1583 & 0.1583 \\
Random mean & 0.0656 & 0.0821 & 0.1252 & 0.1592 & 0.1583 \\
-Random SD (n=5) & 0.0368 & 0.0350 & 0.0262 & 0.0077 & 0.0000 \\

top-mean & 0.0482 & 0.0688 & 0.1239 & 0.1606 & 0.1583 \\
UCB & 0.0482 & 0.0688 & 0.1239 & 0.1606 & 0.1583 \\
\bottomrule
\end{tabular}%
}
\end{table}

\begin{table}[htbp]
\centering
\scriptsize
\caption{GB1 \texttt{2-vs-rest}: top-5\% target recovery by method and budget.}
\label{tab:16-table-h-11-gb1-2-vs-rest-top-5-target-recovery-b}
\resizebox{\linewidth}{!}{%
\begin{tabular}{lrrrrrr}
\toprule
Method & $B=8$ & $B=16$ & $B=32$ & $B=64$ & $B=128$ & $B=256$ \\
\midrule
ITL & 0.0216 & 0.0865 & 0.0457 & 0.0481 & 0.1394 & 0.1803 \\
VTL & 0.1298 & 0.1851 & 0.1370 & 0.1346 & 0.1611 & 0.2524 \\
SampleVariance & 0.0216 & 0.0889 & 0.0457 & 0.0601 & 0.1755 & 0.1779 \\
Random & 0.0346 & 0.0365 & 0.0481 & 0.1034 & 0.1567 & 0.2139 \\
-Random SD (n=5) & 0.0132 & 0.0305 & 0.0275 & 0.0476 & 0.0407 & 0.0090 \\
top-mean & 0.0601 & 0.0433 & 0.0841 & 0.1442 & 0.0361 & 0.2332 \\
UCB & 0.0601 & 0.0938 & 0.0817 & 0.1226 & 0.0577 & 0.2356 \\
\bottomrule
\end{tabular}%
}
\end{table}

\begin{table}[htbp]
\centering
\scriptsize
\caption{GB1 \texttt{3-vs-rest}: top-5\% target recovery by method and budget.}
\label{tab:17-table-h-12-gb1-3-vs-rest-top-5-target-recovery-b}
\resizebox{\linewidth}{!}{%
\begin{tabular}{lrrrrrr}
\toprule
Method & $B=8$ & $B=16$ & $B=32$ & $B=64$ & $B=128$ & $B=256$ \\
\midrule
ITL & 0.1250 & 0.1493 & 0.2465 & 0.2188 & 0.2778 & 0.3194 \\
VTL & 0.0104 & 0.0590 & 0.1562 & 0.2535 & 0.2812 & 0.3542 \\
SampleVariance & 0.1285 & 0.1667 & 0.2465 & 0.2153 & 0.2708 & 0.3056 \\
Random & 0.0938 & 0.0792 & 0.1486 & 0.2118 & 0.2347 & 0.3326 \\
-Random SD (n=5) & 0.0518 & 0.0649 & 0.0828 & 0.0244 & 0.0615 & 0.0531 \\
top-mean & 0.0903 & 0.1424 & 0.1389 & 0.1528 & 0.1632 & 0.2639 \\
UCB & 0.0903 & 0.1424 & 0.1389 & 0.1528 & 0.1632 & 0.2639 \\
\bottomrule
\end{tabular}%
}
\end{table}

The three ProteinGym-MM protocols in Section 4.2 show a consistent method ordering, but with different effect sizes. In canonical $1M\to2M$, ITL/VTL are strongest or near-strongest in both Spearman and top-5\% target recovery, while UCB and top-mean are substantially weaker. This setting provides the most direct evidence that selecting measurements by target-space inference value differs from selecting high-predicted-value or high-local-uncertainty samples within $S$. In the mixed-order pool protocol, the same trend persists under a fold-aware benchmark split: VTL/ITL approach the full-pool reference and recover substantially more top-5\% targets than Random and SampleVariance. In contrast, the Singles-premeasured protocol starts from a stronger posterior, so final-budget method gaps naturally shrink. It is therefore better interpreted as refinement evidence after a high-information warm start rather than as the main evidence for the largest target-aware gains.

The GB1 tables should be interpreted jointly with train-pool size and target order. In \texttt{1-vs-rest}, the accessible pool contains only 28 variants, and the final budget reduces all methods to the same complete-pool fit. Thus, only near-complete budget points provide meaningful method separation; in this small pool, cold top-mean/UCB can also achieve high Spearman. In \texttt{2-vs-rest}, VTL is closest to the Kermut static reference in Spearman and also maintains strong top-5\% recovery; cold top-mean/UCB are clearly above Random at the final budget but remain below VTL. In \texttt{3-vs-rest}, the target is directly $4M$, and low-budget fluctuations are larger, but VTL achieves the highest final-budget Spearman and top-5\% recovery, while top-mean/UCB lag substantially. Overall, results support the main conclusion: target-aware acquisition is clearest when the target space is difficult but still supported by posterior coupling from the accessible pool; in near-exhaustive or lower-order small-pool settings, score-based baselines may be closer to active methods.


\subsection{Additional batch-design and transductive baselines}
\label{app:cr-baselines}

We extend the canonical $1M\to2M$ comparison at $B=64$ on the same
51 ProteinGym-MM assays. All methods use the same accessible and target
sets and the Kermut final predictor. ITL, VTL, conditional SampleVariance,
EPIG, MM-ITL, and CTL use the same Kermut selection posterior; ESM k-center
and target-kernel herding use label-free embedding-based selection.
Random is averaged over five acquisition seeds, while deterministic
methods contribute one selection trajectory per assay.

ESM k-center uses farthest-first selection in normalized ESM embedding
space. Target-kernel herding approximates the target kernel mean while
correcting for coverage already provided by the selected batch. For a
uniform finite target set, EPIG averages $I(y_x;y_a\mid\mathcal D_n)$
over $a\in A$; the latent-target mean-marginal variant, MM-ITL, instead
uses $I(y_x;f_a\mid\mathcal D_n)$. ITL uses information about the joint
latent target vector. CTL sums candidate--target posterior signed
correlations. All posterior-based rules use conditional covariance
updates during batch construction.

\begin{table}[htbp]
\centering
\small
\caption{Canonical ProteinGym-MM at $B=64$. Entries are macro means over
51 assays. Higher Spearman and top-5\% recovery and lower RMSE are better.}
\label{tab:cr-strong-baselines-full}
\begin{tabular}{lrrr}
\toprule
Method & Spearman $\uparrow$ & RMSE $\downarrow$ & Top-5\% $\uparrow$ \\
\midrule
Target-kernel herding & 0.5934 & 0.9005 & 0.3762 \\
ITL                  & 0.5931 & 0.8754 & 0.4243 \\
MM-ITL               & 0.5877 & 0.8900 & 0.3929 \\
EPIG                 & 0.5865 & 0.8797 & 0.4045 \\
CTL                  & 0.5845 & 0.8494 & 0.3779 \\
VTL                  & 0.5756 & 0.8780 & 0.3949 \\
ESM k-center         & 0.4985 & 0.9450 & 0.3168 \\
SampleVariance       & 0.4931 & 0.9986 & 0.3199 \\
Random               & 0.4328 & 1.0175 & 0.2751 \\
\bottomrule
\end{tabular}
\end{table}

The target-aware methods are competitive, with different methods leading
different metrics. Herding has the highest mean Spearman, CTL the lowest
RMSE, and ITL the highest top-5\% recovery. These means do not establish
a statistically resolved ordering among all leading methods.

\paragraph{Why SampleVariance is batch-aware.}
For a fixed Gaussian posterior with independent homoscedastic observation
noise $\tau^2>0$, the Gaussian joint-information objective is
\begin{equation}
G(Q)=\frac12\log\det\left(I+\tau^{-2}\Sigma_n(Q,Q)\right).
\end{equation}
Its greedy marginal gain after selecting $C$ is
\begin{equation}
\Delta G(x\mid C)=\frac12\log\left(1+
\frac{v_n(x\mid C)}{\tau^2}\right),
\end{equation}
where $v_n(x\mid C)$ is the latent candidate variance after conditioning
on noisy observations at $C$. Conditional SampleVariance maximizes this
same variance after each selection. Since the logarithm is strictly
increasing, the two rules produce the same greedy sequence, up to
tie-breaking. Thus, SampleVariance accounts for within-batch redundancy
and is greedy-selection equivalent to Gaussian MaxDet, or the Gaussian
regression specialization of the BatchBALD information objective. This
equivalence does not extend to arbitrary non-Gaussian BatchBALD models.
It also does not imply a general submodularity guarantee for target-aware
ITL or VTL under arbitrary posteriors.

\subsection{Paired statistics and surrogate-quality stratification}
\label{app:cr-singles-statistics}

\paragraph{Paired analysis of singles-premeasured results.}
We reanalyze the original task-level outputs on common feasible tasks.
Random seeds are averaged within each assay-fold task, paired differences
are averaged over folds within assay, and assays receive equal weight.
We obtain pointwise 95\% intervals from 20,000 assay-cluster bootstrap
resamples. These intervals describe variation across assays rather than
treating folds or acquisition seeds as independent proteins; they are not
adjusted for multiple comparisons. At additional budgets
$B=32/64/128/256$, the feasible panels contain 51/51/49/46 assays,
respectively, with five folds per assay.

\begin{table}[htbp]
\centering
\small
\caption{Singles-premeasured: paired Spearman differences with 95\%
assay-cluster bootstrap intervals. SV denotes conditional SampleVariance.}
\label{tab:cr-singles-paired}
\begin{tabular}{llrr}
\toprule
Budget & Method & vs. Random & vs. SV \\
\midrule
32 & ITL & $+0.0278$ [0.0208, 0.0351] & $+0.0344$ [0.0243, 0.0458] \\
32 & VTL & $+0.0311$ [0.0238, 0.0385] & $+0.0378$ [0.0280, 0.0486] \\
64 & ITL & $+0.0204$ [0.0150, 0.0262] & $+0.0240$ [0.0148, 0.0347] \\
64 & VTL & $+0.0284$ [0.0233, 0.0338] & $+0.0320$ [0.0236, 0.0416] \\
128 & ITL & $+0.0211$ [0.0175, 0.0247] & $+0.0218$ [0.0168, 0.0269] \\
128 & VTL & $+0.0176$ [0.0127, 0.0220] & $+0.0183$ [0.0121, 0.0244] \\
256 & ITL & $+0.0106$ [0.0078, 0.0135] & $+0.0076$ [0.0048, 0.0104] \\
256 & VTL & $+0.0037$ [$-0.0012$, 0.0082] & $+0.0007$ [$-0.0040$, 0.0050] \\
\bottomrule
\end{tabular}
\end{table}

All eight early-budget contrasts at $B=32/64$ have positive intervals,
supporting the sample-efficiency interpretation. Differences shrink at
larger budgets, and the $B=256$ VTL contrasts are not statistically
resolved. The static reference is the empirical result of fitting the
same predictor using all protocol-permitted accessible labels; it is
neither a mathematical performance bound nor a reference shared across
different protocols.

\paragraph{Performance when the surrogate is less accurate.}
Separately, in the canonical protocol, we measure surrogate quality by
Kermut's target Spearman after training on all permitted single-mutant
labels. We stratify the 51 assays into three equally sized groups using
this retrospective measure. In the lowest-quality third (17 assays),
ITL gains over Random at $B=32/64/128$ are
$+0.1128/+0.1388/+0.1265$, and VTL gains are
$+0.1230/+0.1098/+0.0973$; all six assay-bootstrap intervals are positive.
At $B=64$, the ITL and VTL gains are $+0.1388$ [0.0600, 0.2257] and
$+0.1098$ [0.0432, 0.1784]. Their corresponding gains over conditional
SampleVariance are $+0.0573$ [$-0.0141$, 0.1259] and $+0.0283$
[$-0.0494$, 0.1078]. Thus, gains over Random are not confined to assays
with high absolute surrogate accuracy, while comparisons against
uncertainty-based acquisition remain less precisely resolved in this
subset. This retrospective stratification is not an experiment-time
criterion for deciding which assays will benefit.

\subsection{Controlled posterior perturbations}
\label{app:cr-misspecification}

We test selection-posterior sensitivity on a canonical panel of 12 assays
at $B=64$. The accessible and target sets and the unperturbed Kermut
final-prediction procedure remain fixed. We change only the selection
posterior, multiplying its observation-noise variance by $0.1$ or $10$,
or replacing the full kernel with sequence-only variants. Candidate and
target labels are not used to choose the perturbation or the batch.
The final predictor is fitted using each selected batch, so downstream
performance differences reflect the different selected measurements.

\begin{table}[htbp]
\centering
\small
\caption{Canonical 12-assay panel, $B=64$: paired Spearman change relative
to the full Kermut selection posterior with noise multiplier 1.
Intervals are pointwise 95\% intervals from 20,000 assay-bootstrap
resamples. ZS denotes the zero-shot mean component.}
\label{tab:cr-perturbations}
\begin{tabular}{llrr}
\toprule
Rule & Selection perturbation & Difference & 95\% interval \\
\midrule
ITL & Noise $0.1\times$ & $-0.0004$ & [$-0.0205$, $+0.0204$] \\
ITL & Noise $10\times$ & $-0.0022$ & [$-0.0126$, $+0.0106$] \\
ITL & Sequence only, no ZS & $+0.0261$ & [$-0.0233$, $+0.0873$] \\
ITL & Sequence only, ZS & $+0.0261$ & [$-0.0233$, $+0.0869$] \\
VTL & Noise $0.1\times$ & $+0.0138$ & [$-0.0147$, $+0.0435$] \\
VTL & Noise $10\times$ & $+0.0438$ & [$-0.0214$, $+0.1209$] \\
VTL & Sequence only, no ZS & $+0.0590$ & [$-0.0011$, $+0.1404$] \\
VTL & Sequence only, ZS & $+0.0590$ & [$-0.0009$, $+0.1399$] \\
\bottomrule
\end{tabular}
\end{table}

Selection changes under these interventions, but downstream Spearman
does not decrease monotonically with perturbation strength. All intervals
in Table~\ref{tab:cr-perturbations} include zero. A noise multiplier changes
the model's assumed measurement noise, rather than adding noise to the
experimental labels; it can change which measurements are selected in
either direction. The experiment measures sensitivity and does not
establish universal robustness to posterior misspecification.

\paragraph{Feedback under perturbed posteriors.}
A separate four-assay panel compares one-shot 64 with $4\times16$
selection at the same total budget. In the multi-round procedure, newly
observed labels are incorporated and the posterior is refitted after each
round. Across six implemented kernel/mean/noise configurations, the mean
feedback-minus-one-shot Spearman differences range from $-0.0163$ to
$+0.0604$ for ITL and $-0.0716$ to $+0.0026$ for VTL. Positive differences
favor feedback. These results show that feedback has no uniform advantage
under posterior perturbations; its effect depends on the posterior
configuration and acquisition objective. The four assays limit the
precision and biological breadth of this comparison. Posterior
information reduction alone does not guarantee monotonic empirical
Spearman after model refitting.

\subsection{Alternative posterior interfaces and additive controls}
\label{app:cr-posterior}

We evaluate alternative interfaces on the singles-premeasured protocol.
The initial observations contain all available single-mutant labels, and
the budget counts additional double-mutant measurements. The panel has
51 assays with five folds each at $B=32/64$; 49 assays remain feasible at
$B=128$. Random seeds are averaged within fold, folds within assay, and
assays are weighted equally.

The Bayesian one-hot model uses an intercept and additive mutation
indicators, without pretrained embeddings, structural features, or
mutation-pair interaction terms. Its prediction for a double mutant is
\begin{equation}
f_{\mathrm{add}}(m_1,m_2)=\beta_0+\beta_{m_1}+\beta_{m_2}.
\end{equation}
We cross the Kermut and one-hot selection posteriors with Kermut and
one-hot final predictors. For a fixed selector and acquisition rule,
both final predictors receive exactly the same selected variants and
labels. This separates predictor comparisons from changes in acquisition.

\begin{table}[htbp]
\centering
\small
\caption{Singles-premeasured alternative-posterior design. Each entry
reports ITL / VTL macro Spearman. Parentheses give the number of feasible
assays; each assay contributes five folds.}
\label{tab:cr-posterior-2x2}
\begin{tabular}{lccc}
\toprule
Selector $\to$ predictor & $B=32$ (51) & $B=64$ (51) & $B=128$ (49) \\
\midrule
Kermut $\to$ Kermut   & 0.7835 / 0.7896 & 0.8231 / 0.8272 & 0.8691 / 0.8696 \\
Kermut $\to$ one-hot  & 0.7377 / 0.7385 & 0.7790 / 0.7817 & 0.8350 / 0.8364 \\
One-hot $\to$ Kermut  & 0.7705 / 0.7759 & 0.8128 / 0.8162 & 0.8564 / 0.8584 \\
One-hot $\to$ one-hot & 0.7462 / 0.7370 & 0.7836 / 0.7795 & 0.8351 / 0.8328 \\
\bottomrule
\end{tabular}
\end{table}

With either selector fixed, Kermut has higher mean Spearman than the
additive predictor in all 12 objective--budget comparisons. Thus, simple
additivity does not fully reproduce the observed prediction performance.
The models also differ in representation and regularization, so their
differences are not isolated measurements of biological epistasis.

\paragraph{A pipeline without pretrained or structural representations.}
When both selection and prediction use the one-hot model, ITL and VTL
improve over Random at every tested budget. This pipeline still uses
experimental single-mutant labels and amino-acid mutation identities;
it does not use Kermut, ESM embeddings, or structural representations.
Table~\ref{tab:cr-onehot-paired} reports paired differences and pointwise
95\% intervals from 20,000 assay-bootstrap resamples. Comparisons against
conditional SampleVariance are smaller and mixed.

\begin{table}[htbp]
\centering
\small
\caption{Fully one-hot pipeline: paired Spearman differences with 95\%
assay-bootstrap intervals. SV denotes conditional SampleVariance.}
\label{tab:cr-onehot-paired}
\begin{tabular}{llrr}
\toprule
Budget & Method & vs. Random & vs. SV \\
\midrule
32 & ITL & $+0.0308$ [0.0217, 0.0411] & $+0.0081$ [$-0.0021$, 0.0187] \\
32 & VTL & $+0.0216$ [0.0148, 0.0289] & $-0.0012$ [$-0.0113$, 0.0096] \\
64 & ITL & $+0.0312$ [0.0232, 0.0399] & $+0.0095$ [0.0019, 0.0182] \\
64 & VTL & $+0.0271$ [0.0201, 0.0348] & $+0.0054$ [$-0.0039$, 0.0162] \\
128 & ITL & $+0.0225$ [0.0164, 0.0289] & $+0.0033$ [0.0004, 0.0065] \\
128 & VTL & $+0.0202$ [0.0156, 0.0252] & $+0.0010$ [$-0.0029$, 0.0052] \\
\bottomrule
\end{tabular}
\end{table}

\paragraph{Bayesian ESM2 last-layer interface.}
We also use a Bayesian linear last layer over frozen ESM2 embeddings as
the selection posterior, keeping the final predictor fixed as Kermut.
Relative to conditional SampleVariance under the same last-layer
posterior, ITL gains $+0.0216/+0.0206/+0.0126$ at $B=32/64/128$, and
VTL gains $+0.0316/+0.0280/+0.0182$. All six corresponding pointwise
95\% assay-bootstrap intervals are positive, and all six comparisons
against Random are also positive. Together, the one-hot and last-layer
experiments support interface flexibility while retaining the requirement
for informative candidate--target covariance.

\section{Additional ProteinGym-MM Transfer Protocols}
\label{appi}

The ProteinGym-MM experiments in Section 4.2 focus on three mutation-space protocols: canonical $1M\to2M$, the mixed-order pool protocol, and the Singles-premeasured protocol. This appendix adds two  auxiliary protocols to examine how mixed-order accessible pool behaves under other target definitions. We start from an empty initial labeled set, $L_0=\varnothing$, settings are $S=1M_{\mathrm{train}}\cup2M_{\mathrm{train}} \to A=1M_{\mathrm{test}}$ and $S=1M\cup2M \to A= \ge 3M$.

\begin{figure}[htbp]
\centering
\includegraphics[width=0.7\linewidth]{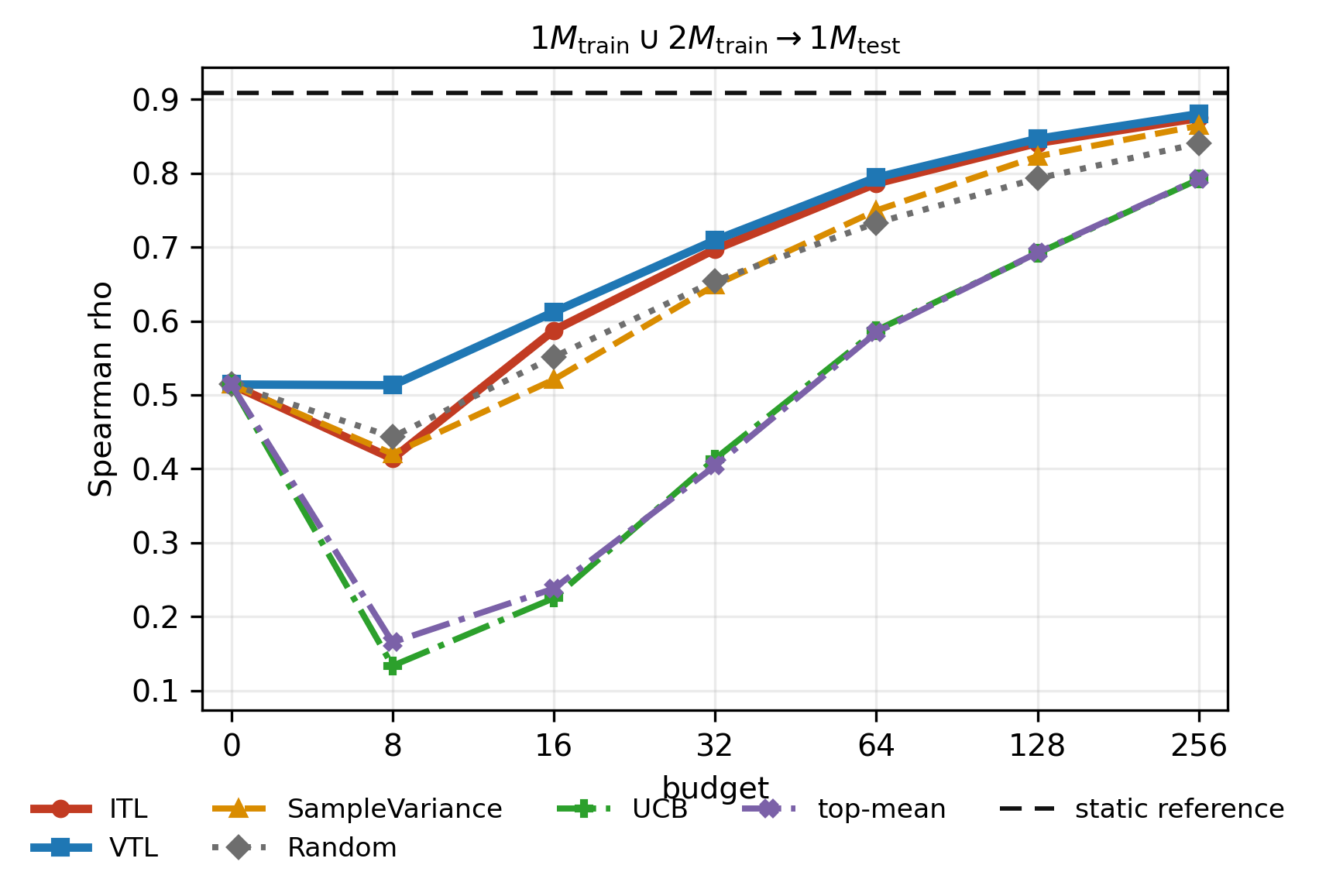}
\caption{Auxiliary ProteinGym-MM mutation-space protocols.  The protocol uses $L_0=\varnothing$ and $S=1M_{\mathrm{train}}\cup2M_{\mathrm{train}}$, and evaluate target-side $1M$.}
\label{fig:auxiliary-protocols}
\end{figure}

\begin{figure}[htbp]
\centering
\includegraphics[width=\linewidth]{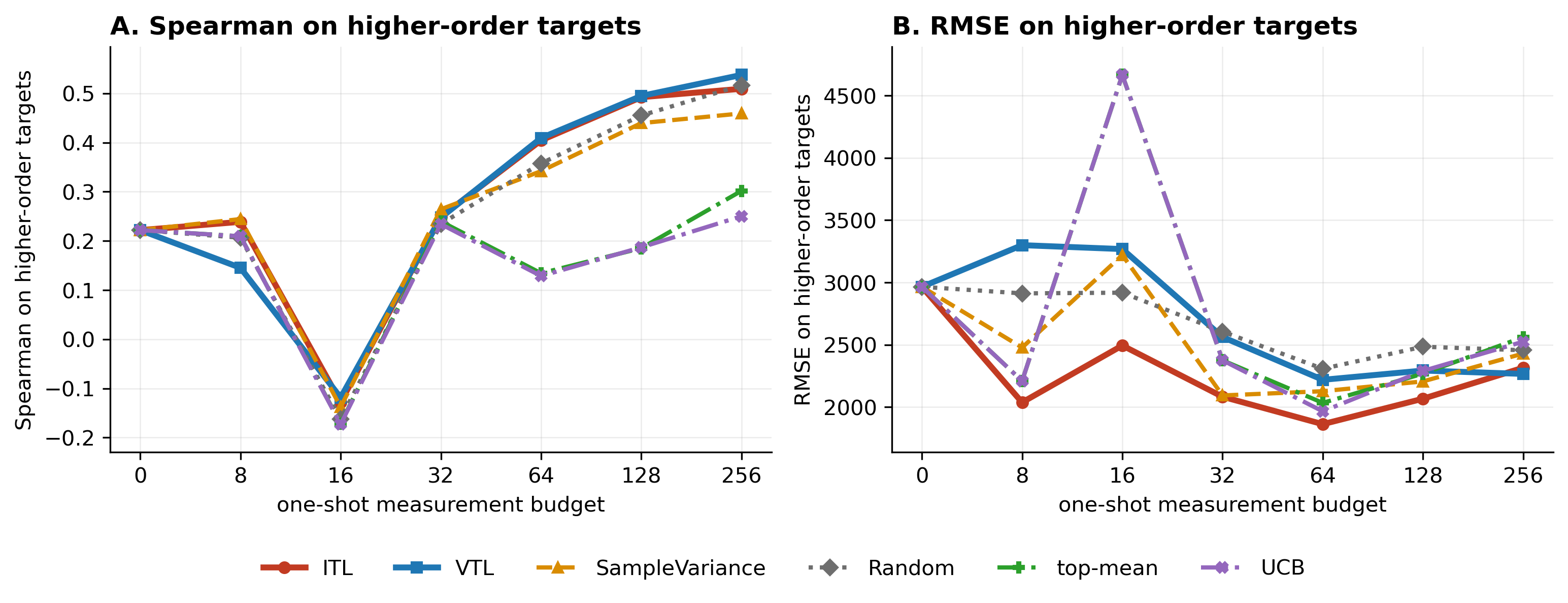}
\caption{Higher-order cold-start ProteinGym-MM extension on 11 assays. The protocol uses $L_0=\varnothing$, $S=1M\cup2M$, and $A=\geq 3M$.}
\label{fig:ho11-empty-l0-higher-order}
\end{figure}

In the target-side $1M$ protocol, VTL and ITL lead throughout the medium-to-high budget range. At the final budget, VTL reaches \texttt{0.8802}, above Random at \texttt{0.8403} and SampleVariance at \texttt{0.8644}, reducing the gap to the static reference to \texttt{0.0283}. This result indicates that even when the target set changes from $2M$ to $1M$, posterior coupling can still help measurements from the mixed-order pool serve target-side prediction more effectively. At the same time, top-mean and UCB are substantially below ITL/VTL in this protocol, suggesting that ranking within the measurable pool by posterior mean or mean-plus-uncertainty is still insufficient to replace target-oriented acquisition.

We also evaluated a stricter sparse ProteinGym-MM setting in which no variants are initially labeled, both single and double mutants are queryable, and variants with three or more substitutions form the target space:
$L_0=\varnothing$, $S=1M\cup2M$, and $A=\geq 3M$.
This protocol asks whether a low-order accessible pool can support higher-order target prediction from a cold start.
It is substantially more heterogeneous than the canonical $1M\to2M$ setting because target sizes, low-order pool sizes, and higher-order mutation distributions vary strongly across assays.
We therefore use it as scalability and boundary evidence, rather than as a primary headline result.

In this higher-order cold-start setting, VTL gives the best final Spearman, RMSE, and top-fraction recovery.
Compared with Random, VTL improves final Spearman by $0.0227$, reduces RMSE by about $155$, and improves top-5\% target recovery by $0.0361$.
ITL is close to Random in final Spearman but gives slightly better RMSE and top-5\% recovery.
By contrast, score-based baselines are substantially weaker: top-mean and UCB reach only $0.3120$ and $0.2685$ final Spearman, respectively. This indicates that selecting low-order variants by their own posterior mean or mean-plus-uncertainty is not a reliable substitute for target-aware measurement value when the goal is prediction over higher-order combinatorial targets.

\section{Target-Uncertainty Diagnostics in Canonical $1M\to2M$}
\label{app:target-uncertainty-diagnostics}

As an additional diagnostic in the canonical $1M\to2M$ setting, we ask whether ITL/VTL improve target prediction while also reducing posterior uncertainty on $A=2M$.

\begin{figure}[htbp]
\centering
\includegraphics[width=\linewidth]{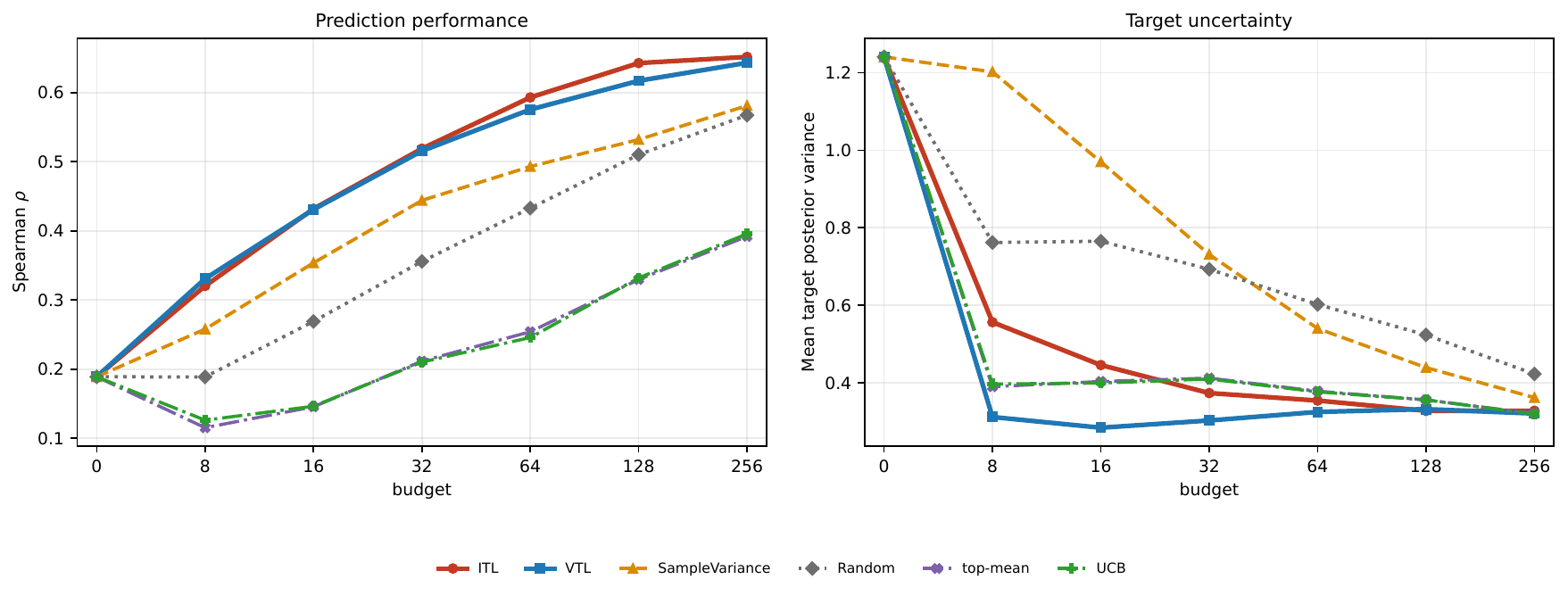}
\caption{Target prediction and posterior uncertainty in canonical ProteinGym-MM $1M\to2M$. The left panel reports target Spearman $\rho$ on $A=2M$ as a function of budget. The right panel reports mean target posterior variance, $|A|^{-1}\operatorname{tr}\Sigma_n(A,A)$, after updating the posterior with the measured singles acquired so far. Curves are macro-averaged over the 51 assays.}
\label{fig:target-uncertainty-diagnostics}
\end{figure}

We define target uncertainty as the average marginal variance of the current posterior on the target set, $|A|^{-1}\operatorname{tr}\Sigma_n(A,A)$. At each budget, we update the posterior with the measured singles acquired so far, compute this quantity on $A=2M$, and macro-average it over the 51 assays. Thus, the right panel of Figure~\ref{fig:target-uncertainty-diagnostics} measures the model's internal residual uncertainty over target doubles, rather than final prediction error itself.

Figure~\ref{fig:target-uncertainty-diagnostics} shows that ITL/VTL improve target ranking performance while also reducing posterior uncertainty on $A$ more quickly. At $B=128$, the Spearman values of ITL/VTL are \texttt{0.6428\ /\ 0.6172}, above Random at \texttt{0.5103}; their corresponding mean target posterior variances are \texttt{0.3281\ /\ 0.3324}, also below Random at \texttt{0.5237}. At the same time, top-mean/UCB also reduce posterior variance to \texttt{0.3564\ /\ 0.3561}, but their Spearman values are only \texttt{0.3297\ /\ 0.3320}; SampleVariance reaches only \texttt{0.5321} Spearman. Thus, simply reducing posterior uncertainty or selecting high-scoring singles is insufficient to explain target-transfer gains. What matters is whether the measurements actually support ranking improvement.

\begin{table}[htbp]
\centering
\scriptsize
\caption{Higher-order cold-start extension: final-budget summary. Metrics are macro-averaged over each assay's largest budget at which all six methods are available.}
\label{tab:ho11-empty-l0-final-summary}
\begin{tabular}{lrrrr}
\toprule
Method & Spearman $\uparrow$ & RMSE $\downarrow$ & top-5\% $\uparrow$ & top-10\% $\uparrow$ \\
\midrule
ITL & 0.4797 & 1894.52 & 0.2559 & 0.3224 \\
VTL & 0.5017 & 1853.86 & 0.2764 & 0.3457 \\
SampleVariance & 0.4315 & 1987.73 & 0.2215 & 0.2897 \\
Random & 0.4790 & 2009.15 & 0.2403 & 0.3199 \\
top-mean & 0.3120 & 2095.16 & 0.1698 & 0.2421 \\
UCB & 0.2685 & 2065.49 & 0.1708 & 0.2374 \\
\bottomrule
\end{tabular}
\end{table}

\section{Reproducibility Statement}

\textbf{Experimental details.}
The main experimental protocols are specified using the $L_0,S,A$ notation in Section~\ref{35} and instantiated in Section~\ref{41}. Sections~\ref{42}-\ref{44} describe the main ProteinGym-MM and GB1 protocols, baselines, budgets, evaluation metrics. Appendix~\ref{appg} reports budget coverage and Random-equivalent sample efficiency, Appendix\ref{apph} provides full budget-performance tables, and Appendix~\ref{appi} reports additional ProteinGym-MM transfer protocols. Random baselines are averaged over five seeds unless otherwise stated.

\textbf{Data and code access.}
The experiments use ProteinGym-derived multi-mutant assays and the GB1 complete landscape derived from the FLIP context. ProteinGym benchmark materials are available from the project website \url{https://www.proteingym.org/} and repository \url{https://github.com/OATML-Markslab/ProteinGym}; FLIP scripts and data are available at \url{https://benchmark.protein.properties}; Transductive Active Learning related code are available at \url{https://github.com/jonhue/transductive-active-learning}. The repository contains the scripts, configurations, exported figure data, and result summaries used to produce the reported tables and figures. We will release the TALK code, processed experiment manifests, and reproduction instructions with the camera-ready material.

\textbf{Compute environment and runtime.}
All reported experiments were run on NVIDIA RTX 4090D GPUs. As a representative runtime estimate, A ProteinGym-MM-51 run contains 306 completed runs, corresponding to 51 assays, two acquisition methods, and three seeds. Using the recorded start time and aggregate-output timestamp, this run took approximately 2.01 hours, or about 2.36 minutes per assay for the full two-method, three-seed setting. This number is hardware- and protocol-specific; runtime varies with assay size, target-space size, posterior backend, budget schedule, and the number of seeds and acquisition rules.

\section{Broader impacts}
TALK is intended to make fixed-budget protein variant experiments more sample-efficient by selecting measurements that improve prediction on a specified target space. Positive impacts include reducing experimental cost and material use, helping smaller laboratories make better use of limited assay budgets, and supporting beneficial protein-engineering applications such as enzyme optimization, protein stability analysis, and therapeutic candidate prioritization. At the same time, any method that improves the efficiency of protein variant selection can have dual-use implications: in inappropriate settings, target-aware experimental design could reduce the cost of optimizing harmful or unsafe biological functions. The method may also amplify inequities if only well-resourced groups can combine active selection with high-throughput assays.

\section{Safeguards}
This work does not release a pretrained generative biological model, a scraped biological dataset, pathogenic sequence collections, wet-lab optimized variants, or an autonomous design pipeline. The experiments are retrospective evaluations on established benchmark-style protein variant data, and the released artifacts are limited to the manuscript, evaluation code, aggregate results, and non-sensitive figure data. Responsible use of TALK in prospective wet-lab settings should remain subject to institutional biosafety review, DNA synthesis screening, application-specific risk assessment, and restrictions on pathogen-, toxin-, or virulence-related optimization tasks.


\newpage
\section*{NeurIPS Paper Checklist}


\begin{enumerate}

\item {\bf Claims}
    \item[] Question: Do the main claims made in the abstract and introduction accurately reflect the paper's contributions and scope?
    \item[] Answer: \answerYes{} 
    \item[] Justification: The abstract and introduction accurately describe TALK as a fixed-budget target-aware measurement-design framework and state claims that are supported by the method and retrospective experiments on ProteinGym-MM and GB1 in Section 4. The claims are scoped to the evaluated protein variant settings and do not assert general guarantees across all protein design tasks or predictor classes.
    \item[] Guidelines:
    \begin{itemize}
        \item The answer \answerNA{} means that the abstract and introduction do not include the claims made in the paper.
        \item The abstract and/or introduction should clearly state the claims made, including the contributions made in the paper and important assumptions and limitations. A \answerNo{} or \answerNA{} answer to this question will not be perceived well by the reviewers. 
        \item The claims made should match theoretical and experimental results, and reflect how much the results can be expected to generalize to other settings. 
        \item It is fine to include aspirational goals as motivation as long as it is clear that these goals are not attained by the paper. 
    \end{itemize}

\item {\bf Limitations}
    \item[] Question: Does the paper discuss the limitations of the work performed by the authors?
    \item[] Answer: \answerYes{} 
    \item[] Justification: In 5. Discussion.
    \item[] Guidelines:
    \begin{itemize}
        \item The answer \answerNA{} means that the paper has no limitation while the answer \answerNo{} means that the paper has limitations, but those are not discussed in the paper. 
        \item The authors are encouraged to create a separate ``Limitations'' section in their paper.
        \item The paper should point out any strong assumptions and how robust the results are to violations of these assumptions (e.g., independence assumptions, noiseless settings, model well-specification, asymptotic approximations only holding locally). The authors should reflect on how these assumptions might be violated in practice and what the implications would be.
        \item The authors should reflect on the scope of the claims made, e.g., if the approach was only tested on a few datasets or with a few runs. In general, empirical results often depend on implicit assumptions, which should be articulated.
        \item The authors should reflect on the factors that influence the performance of the approach. For example, a facial recognition algorithm may perform poorly when image resolution is low or images are taken in low lighting. Or a speech-to-text system might not be used reliably to provide closed captions for online lectures because it fails to handle technical jargon.
        \item The authors should discuss the computational efficiency of the proposed algorithms and how they scale with dataset size.
        \item If applicable, the authors should discuss possible limitations of their approach to address problems of privacy and fairness.
        \item While the authors might fear that complete honesty about limitations might be used by reviewers as grounds for rejection, a worse outcome might be that reviewers discover limitations that aren't acknowledged in the paper. The authors should use their best judgment and recognize that individual actions in favor of transparency play an important role in developing norms that preserve the integrity of the community. Reviewers will be specifically instructed to not penalize honesty concerning limitations.
    \end{itemize}

\item {\bf Theory assumptions and proofs}
    \item[] Question: For each theoretical result, does the paper provide the full set of assumptions and a complete (and correct) proof?
    \item[] Answer: \answerYes{} 
    \item[] Justification: We give necessary demonstration in 3.Method and corresponding Appendices mentioned in Method. We have not repeatedly presented others' proofs.
    \item[] Guidelines:
    \begin{itemize}
        \item The answer \answerNA{} means that the paper does not include theoretical results. 
        \item All the theorems, formulas, and proofs in the paper should be numbered and cross-referenced.
        \item All assumptions should be clearly stated or referenced in the statement of any theorems.
        \item The proofs can either appear in the main paper or the supplemental material, but if they appear in the supplemental material, the authors are encouraged to provide a short proof sketch to provide intuition. 
        \item Inversely, any informal proof provided in the core of the paper should be complemented by formal proofs provided in appendix or supplemental material.
        \item Theorems and Lemmas that the proof relies upon should be properly referenced. 
    \end{itemize}

    \item {\bf Experimental result reproducibility}
    \item[] Question: Does the paper fully disclose all the information needed to reproduce the main experimental results of the paper to the extent that it affects the main claims and/or conclusions of the paper (regardless of whether the code and data are provided or not)?
    \item[] Answer: \answerYes{}{} 
    \item[] Justification: Appendix. Reproducibility
    \item[] Guidelines:
    \begin{itemize}
        \item The answer \answerNA{} means that the paper does not include experiments.
        \item If the paper includes experiments, a \answerNo{} answer to this question will not be perceived well by the reviewers: Making the paper reproducible is important, regardless of whether the code and data are provided or not.
        \item If the contribution is a dataset and\slash or model, the authors should describe the steps taken to make their results reproducible or verifiable. 
        \item Depending on the contribution, reproducibility can be accomplished in various ways. For example, if the contribution is a novel architecture, describing the architecture fully might suffice, or if the contribution is a specific model and empirical evaluation, it may be necessary to either make it possible for others to replicate the model with the same dataset, or provide access to the model. In general. releasing code and data is often one good way to accomplish this, but reproducibility can also be provided via detailed instructions for how to replicate the results, access to a hosted model (e.g., in the case of a large language model), releasing of a model checkpoint, or other means that are appropriate to the research performed.
        \item While NeurIPS does not require releasing code, the conference does require all submissions to provide some reasonable avenue for reproducibility, which may depend on the nature of the contribution. For example
        \begin{enumerate}
            \item If the contribution is primarily a new algorithm, the paper should make it clear how to reproduce that algorithm.
            \item If the contribution is primarily a new model architecture, the paper should describe the architecture clearly and fully.
            \item If the contribution is a new model (e.g., a large language model), then there should either be a way to access this model for reproducing the results or a way to reproduce the model (e.g., with an open-source dataset or instructions for how to construct the dataset).
            \item We recognize that reproducibility may be tricky in some cases, in which case authors are welcome to describe the particular way they provide for reproducibility. In the case of closed-source models, it may be that access to the model is limited in some way (e.g., to registered users), but it should be possible for other researchers to have some path to reproducing or verifying the results.
        \end{enumerate}
    \end{itemize}

\item {\bf Open access to data and code}
    \item[] Question: Does the paper provide open access to the data and code, with sufficient instructions to faithfully reproduce the main experimental results, as described in supplemental material?
    \item[] Answer: \answerYes{} 
    \item[] Justification: We will make the code files public before the article is published.
    \item[] Guidelines:
    \begin{itemize}
        \item The answer \answerNA{} means that paper does not include experiments requiring code.
        \item Please see the NeurIPS code and data submission guidelines (\url{https://neurips.cc/public/guides/CodeSubmissionPolicy}) for more details.
        \item While we encourage the release of code and data, we understand that this might not be possible, so \answerNo{} is an acceptable answer. Papers cannot be rejected simply for not including code, unless this is central to the contribution (e.g., for a new open-source benchmark).
        \item The instructions should contain the exact command and environment needed to run to reproduce the results. See the NeurIPS code and data submission guidelines (\url{https://neurips.cc/public/guides/CodeSubmissionPolicy}) for more details.
        \item The authors should provide instructions on data access and preparation, including how to access the raw data, preprocessed data, intermediate data, and generated data, etc.
        \item The authors should provide scripts to reproduce all experimental results for the new proposed method and baselines. If only a subset of experiments are reproducible, they should state which ones are omitted from the script and why.
        \item At submission time, to preserve anonymity, the authors should release anonymized versions (if applicable).
        \item Providing as much information as possible in supplemental material (appended to the paper) is recommended, but including URLs to data and code is permitted.
    \end{itemize}

\item {\bf Experimental setting/details}
    \item[] Question: Does the paper specify all the training and test details (e.g., data splits, hyperparameters, how they were chosen, type of optimizer) necessary to understand the results?
    \item[] Answer: \answerYes{}{} 
    \item[] Justification: Appendix. Reproducibility
    \item[] Guidelines:
    \begin{itemize}
        \item The answer \answerNA{} means that the paper does not include experiments.
        \item The experimental setting should be presented in the core of the paper to a level of detail that is necessary to appreciate the results and make sense of them.
        \item The full details can be provided either with the code, in appendix, or as supplemental material.
    \end{itemize}

\item {\bf Experiment statistical significance}
    \item[] Question: Does the paper report error bars suitably and correctly defined or other appropriate information about the statistical significance of the experiments?
    \item[] Answer: \answerYes{} 
    \item[] Justification: We provide Random SD (n=5) for all experiments with randomness, in Appendix H.
    \item[] Guidelines:
    \begin{itemize}
        \item The answer \answerNA{} means that the paper does not include experiments.
        \item The authors should answer \answerYes{} if the results are accompanied by error bars, confidence intervals, or statistical significance tests, at least for the experiments that support the main claims of the paper.
        \item The factors of variability that the error bars are capturing should be clearly stated (for example, train/test split, initialization, random drawing of some parameter, or overall run with given experimental conditions).
        \item The method for calculating the error bars should be explained (closed form formula, call to a library function, bootstrap, etc.)
        \item The assumptions made should be given (e.g., Normally distributed errors).
        \item It should be clear whether the error bar is the standard deviation or the standard error of the mean.
        \item It is OK to report 1-sigma error bars, but one should state it. The authors should preferably report a 2-sigma error bar than state that they have a 96\% CI, if the hypothesis of Normality of errors is not verified.
        \item For asymmetric distributions, the authors should be careful not to show in tables or figures symmetric error bars that would yield results that are out of range (e.g., negative error rates).
        \item If error bars are reported in tables or plots, the authors should explain in the text how they were calculated and reference the corresponding figures or tables in the text.
    \end{itemize}

\item {\bf Experiments compute resources}
    \item[] Question: For each experiment, does the paper provide sufficient information on the computer resources (type of compute workers, memory, time of execution) needed to reproduce the experiments?
    \item[] Answer: \answerYes{}{} 
    \item[] Justification: Appendix. Reproducibility
    \item[] Guidelines:
    \begin{itemize}
        \item The answer \answerNA{} means that the paper does not include experiments.
        \item The paper should indicate the type of compute workers CPU or GPU, internal cluster, or cloud provider, including relevant memory and storage.
        \item The paper should provide the amount of compute required for each of the individual experimental runs as well as estimate the total compute. 
        \item The paper should disclose whether the full research project required more compute than the experiments reported in the paper (e.g., preliminary or failed experiments that didn't make it into the paper). 
    \end{itemize}
    
\item {\bf Code of ethics}
    \item[] Question: Does the research conducted in the paper conform, in every respect, with the NeurIPS Code of Ethics \url{https://neurips.cc/public/EthicsGuidelines}?
    \item[] Answer: \answerYes{} 
    \item[] Justification: We follow the NeurIPS Code of Ethics.
    \item[] Guidelines:
    \begin{itemize}
        \item The answer \answerNA{} means that the authors have not reviewed the NeurIPS Code of Ethics.
        \item If the authors answer \answerNo, they should explain the special circumstances that require a deviation from the Code of Ethics.
        \item The authors should make sure to preserve anonymity (e.g., if there is a special consideration due to laws or regulations in their jurisdiction).
    \end{itemize}

\item {\bf Broader impacts}
    \item[] Question: Does the paper discuss both potential positive societal impacts and negative societal impacts of the work performed?
    \item[] Answer: \answerYes{} 
    \item[] Justification: Appendix. Broad Impacts.
    \item[] Guidelines:
    \begin{itemize}
        \item The answer \answerNA{} means that there is no societal impact of the work performed.
        \item If the authors answer \answerNA{} or \answerNo, they should explain why their work has no societal impact or why the paper does not address societal impact.
        \item Examples of negative societal impacts include potential malicious or unintended uses (e.g., disinformation, generating fake profiles, surveillance), fairness considerations (e.g., deployment of technologies that could make decisions that unfairly impact specific groups), privacy considerations, and security considerations.
        \item The conference expects that many papers will be foundational research and not tied to particular applications, let alone deployments. However, if there is a direct path to any negative applications, the authors should point it out. For example, it is legitimate to point out that an improvement in the quality of generative models could be used to generate Deepfakes for disinformation. On the other hand, it is not needed to point out that a generic algorithm for optimizing neural networks could enable people to train models that generate Deepfakes faster.
        \item The authors should consider possible harms that could arise when the technology is being used as intended and functioning correctly, harms that could arise when the technology is being used as intended but gives incorrect results, and harms following from (intentional or unintentional) misuse of the technology.
        \item If there are negative societal impacts, the authors could also discuss possible mitigation strategies (e.g., gated release of models, providing defenses in addition to attacks, mechanisms for monitoring misuse, mechanisms to monitor how a system learns from feedback over time, improving the efficiency and accessibility of ML).
    \end{itemize}
    
\item {\bf Safeguards}
    \item[] Question: Does the paper describe safeguards that have been put in place for responsible release of data or models that have a high risk for misuse (e.g., pre-trained language models, image generators, or scraped datasets)?
    \item[] Answer: \answerYes{} 
    \item[] Justification: Appendix. Safeguards.
    \item[] Guidelines:
    \begin{itemize}
        \item The answer \answerNA{} means that the paper poses no such risks.
        \item Released models that have a high risk for misuse or dual-use should be released with necessary safeguards to allow for controlled use of the model, for example by requiring that users adhere to usage guidelines or restrictions to access the model or implementing safety filters. 
        \item Datasets that have been scraped from the Internet could pose safety risks. The authors should describe how they avoided releasing unsafe images.
        \item We recognize that providing effective safeguards is challenging, and many papers do not require this, but we encourage authors to take this into account and make a best faith effort.
    \end{itemize}

\item {\bf Licenses for existing assets}
    \item[] Question: Are the creators or original owners of assets (e.g., code, data, models), used in the paper, properly credited and are the license and terms of use explicitly mentioned and properly respected?
    \item[] Answer: \answerYes{}{} 
    \item[] Justification: All existing assets used in this work, including public datasets, pretrained models, and baseline code, are publicly available for research use. We credit the original creators through citations and explicitly mention the corresponding sources where they are introduced. We have checked and complied with their stated licenses and terms of use.
    \item[] Guidelines:
    \begin{itemize}
        \item The answer \answerNA{} means that the paper does not use existing assets.
        \item The authors should cite the original paper that produced the code package or dataset.
        \item The authors should state which version of the asset is used and, if possible, include a URL.
        \item The name of the license (e.g., CC-BY 4.0) should be included for each asset.
        \item For scraped data from a particular source (e.g., website), the copyright and terms of service of that source should be provided.
        \item If assets are released, the license, copyright information, and terms of use in the package should be provided. For popular datasets, \url{paperswithcode.com/datasets} has curated licenses for some datasets. Their licensing guide can help determine the license of a dataset.
        \item For existing datasets that are re-packaged, both the original license and the license of the derived asset (if it has changed) should be provided.
        \item If this information is not available online, the authors are encouraged to reach out to the asset's creators.
    \end{itemize}

\item {\bf New assets}
    \item[] Question: Are new assets introduced in the paper well documented and is the documentation provided alongside the assets?
    \item[] Answer: \answerNA{} 
    \item[] Justification: \answerNA{}
    \item[] Guidelines:
    \begin{itemize}
        \item The answer \answerNA{} means that the paper does not release new assets.
        \item Researchers should communicate the details of the dataset\slash code\slash model as part of their submissions via structured templates. This includes details about training, license, limitations, etc. 
        \item The paper should discuss whether and how consent was obtained from people whose asset is used.
        \item At submission time, remember to anonymize your assets (if applicable). You can either create an anonymized URL or include an anonymized zip file.
    \end{itemize}

\item {\bf Crowdsourcing and research with human subjects}
    \item[] Question: For crowdsourcing experiments and research with human subjects, does the paper include the full text of instructions given to participants and screenshots, if applicable, as well as details about compensation (if any)? 
    \item[] Answer: \answerNA{} 
    \item[] Justification: \answerNA{}
    \item[] Guidelines:
    \begin{itemize}
        \item The answer \answerNA{} means that the paper does not involve crowdsourcing nor research with human subjects.
        \item Including this information in the supplemental material is fine, but if the main contribution of the paper involves human subjects, then as much detail as possible should be included in the main paper. 
        \item According to the NeurIPS Code of Ethics, workers involved in data collection, curation, or other labor should be paid at least the minimum wage in the country of the data collector. 
    \end{itemize}

\item {\bf Institutional review board (IRB) approvals or equivalent for research with human subjects}
    \item[] Question: Does the paper describe potential risks incurred by study participants, whether such risks were disclosed to the subjects, and whether Institutional Review Board (IRB) approvals (or an equivalent approval/review based on the requirements of your country or institution) were obtained?
    \item[] Answer: \answerNA{} 
    \item[] Justification: \answerNA{}
    \item[] Guidelines:
    \begin{itemize}
        \item The answer \answerNA{} means that the paper does not involve crowdsourcing nor research with human subjects.
        \item Depending on the country in which research is conducted, IRB approval (or equivalent) may be required for any human subjects research. If you obtained IRB approval, you should clearly state this in the paper. 
        \item We recognize that the procedures for this may vary significantly between institutions and locations, and we expect authors to adhere to the NeurIPS Code of Ethics and the guidelines for their institution. 
        \item For initial submissions, do not include any information that would break anonymity (if applicable), such as the institution conducting the review.
    \end{itemize}

\item {\bf Declaration of LLM usage}
    \item[] Question: Does the paper describe the usage of LLMs if it is an important, original, or non-standard component of the core methods in this research? Note that if the LLM is used only for writing, editing, or formatting purposes and does \emph{not} impact the core methodology, scientific rigor, or originality of the research, declaration is not required.
    \item[] Answer: \answerNA{} 
    \item[] Justification: \answerNA{}
    \item[] Guidelines:
    \begin{itemize}
        \item The answer \answerNA{} means that the core method development in this research does not involve LLMs as any important, original, or non-standard components.
        \item Please refer to our LLM policy in the NeurIPS handbook for what should or should not be described.
    \end{itemize}

\end{enumerate}

\end{document}